\documentstyle[pra,aps,eqsecnum,epsfig]{revtex}

\def\ut#1{\lower1.2ex\hbox{$\mathchar"3218$}\mkern -14mu%
          \hbox to 2ex{\hss$#1$\hss}}

\begin{document}
\draft

\title{Finite temperature hydrodynamic modes of trapped quantum gases}
\author{Andr\'as Csord\'as}
\address{Research Group for Statistical Physics of the Hungarian
         Academy of Sciences\\
         P\'azm\'any P\'eter S\'et\'any 1/A\\
         1117 Budapest, Hungary}
\author{Robert Graham}
\address{Fachbereich Physik, Universit\"at-GH, \\
         45117 Essen, Germany}
\maketitle
\begin{abstract}
       The hydrodynamic equations of an ideal fluid formed by a dilute quantum
gas in a parabolic trapping  potential are studied analytically and
numerically. Due to the appearance of internal modes in the fluid stratified  
by the trapping potential, the spectrum of low-lying modes is found to be
dense in the high-temperature limit, with an infinitely degenerate set of
zero-frequency modes. The spectrum for Bose-fluids and Fermi-fluids is
obtained and discussed.
       \end{abstract}

\vskip2pc
\section{Introduction}\label{sec:1}

The successful trapping of dilute Bose- and Fermi-gases in magnetic traps and  
their subsequent cooling to temperatures below quantum-degeneracy  
\cite{1,2,3} has made
the study of their hydrodynamics a subject of high current interest. The basic  
hydrodynamic equations of the fluid formed by such gases in local
thermodynamic equilibrium are well known. In the limit where the fluid can be  
considered ideal they are the continuity equation for the mass-density $\rho$  
and velocity field $\vec{u}$
\begin{eqnarray}
\frac{\partial\rho}{\partial t}+\vec{\nabla}\cdot\rho\vec{u}=0,
\label{eq:1.1}
\end{eqnarray}
the Euler equation for the velocity field
\begin{eqnarray}
\frac{\partial\vec{u}}{\partial t}+\vec{u}\cdot\vec{\nabla}\vec{u}
    = -\frac{1}{\rho}\vec{\nabla}P+\vec{f}(\vec{x}),
\label{eq:1.2}
\end{eqnarray}
with the external force per unit mass
      $\vec{f}=-\frac{1}{m}\vec{\nabla}V(\vec{x})$
and the pressure $P$ related to the internal energy density $\varepsilon$ by  
$P=\frac{2}{3}\varepsilon$. It satisfies
\begin{eqnarray}
\frac{\partial P}{\partial t}+\vec{u}\cdot\vec{\nabla} P
     = -\frac{5}{3}\left[\vec{\nabla}(\vec{u}P)-\rho\vec{u}\cdot\vec{f}\right]
\label{eq:1.3}
\end{eqnarray}
The thermodynamic equilibrium distributions of the density $\rho_0(\vec{x})$  
and pressure
     $P_0(\vec{x})$ with $\vec{\nabla}P_0=\rho_0\vec{f}$ are given by the  
ideal quantum-gas expressions at
constant temperature $\beta=1/k_BT$ and chemical potential $\mu$
\begin{eqnarray}
\rho_0(\vec{x})
     &&= m\int\frac{d^3k}{(2\pi)^3} f_{\mp}
      (\vec{k},\vec{x})
\label{eq:1.4}\\
 P_0 (\vec{x})
     &&= \frac{2}{3}\int\frac{d^3k}{(2\pi)^3}
       \frac{\hbar^2k^2}{2m}f_{\mp}(\vec{k},\vec{x}),
\label{eq:1.5}
\end{eqnarray}
with the single-particle distribution
\begin{eqnarray}
f_{\mp}\left(\vec{k},\vec{x}\right)
     = \frac{1}{e^{\beta
       \left(\frac{\hbar^2k^2}{2m}+V(\vec{x})-\mu\right)}\mp 1}
\label{eq:1.6}
\end{eqnarray}
of the Bose-Einstein (upper sign) or Fermi-Dirac (lower sign) form.

The derivation of these equations from the Boltzmann-equation is well-known,  
see \cite{BK}.  In recent years a number of papers have already been devoted  
to the study of solutions of these equations linearized around the equilibrium  
state in parabolic traps. Griffin, Wu and Stringari \cite{GWS} derived a
closed equation for the velocity fluctuations and gave explicit solutions for  
surface waves of a Bose gas in an isotropic trap and, for a classical gas,
also in the axially symmetric anisotropic trap.  In the latter case they gave  
also  solutions for modes corresponding to irrotational flow. A further  
studies of the
hydrodynamic regime of a trapped bose-gases  was presented in \cite{Kagan}.   
Fermi-gases were considered by Bruun and Clark
\cite{BC}. Besides considering the low temperature limit for the degenerate
Fermi gas, these authors gave an analytical solution for the mode spectrum in  
an isotropic
trap in the high-temperature limit and identified one branch of the dispersion  
relation as `internal waves' driven by the inhomogeneous trap potential. This  
is a point which we intend to examine further in the present paper. Amoruso
et al. \cite{A} also derived special solutions to the linearized hydrodynamic  
equations for the low-temperature limit of the degenerate Fermi gas. In a
recent paper \cite{CG} we have also studied this low-temperature regime for  
Fermi gases
and gave solutions for the completely anisotropic parabolic trap. The present  
paper will therefore concentrate on temperatures of the order of the
degeneracy temperature or above. In a number of papers effects beyond the
scope of  eqs.(\ref{eq:1.1}) - (\ref{eq:1.6}) were also considered. Vichi and  
Stringari \cite{VS} considered the effects of mean-fields due to interactions  
on the collective oscillations of Fermi gases in a trap, while Pethick, Smith  
and collaborators\cite{KPS1}, \cite{KPS2}, \cite{APS}, Vichi \cite{V} and
Gu\'ery-Odelin et al. \cite{GO}
discussed the collisional damping of collective modes in Bose gases and Fermi  
gases respectively. In \cite{KPS1,KPS2} a simple interpolation formula was  
proposed
between the mode-frequencies $\omega_c$ in the collisionless regime
and the hydrodynamic regime, $\omega_h$, of the form
\begin{equation}
\omega^2=\omega_c^2+\frac{\omega_h^2-\omega_c^2}{1-i\omega\tau}
\label{int}
\end{equation}
where
\begin{equation}
\tau^{-1}=\frac{8\pi a^2}{m}<\rho_0 v>
\label{tau}\end{equation}
is the mean collision rate. This description was further examined in  
ref.\cite{GO}. Eq.(\ref{int}) is based on general considerations
of non-equilibrium thermodynamics \cite{LL}.
Damping of the hydrodynamics in Bose gases was also studied by Griffin and
collaborators in a series of papers, see \cite{GriffinHyd} where further  
references can be found.

In the present paper we will not be concerned with damping effects. Instead
 our goal in the present paper is to study further the
collision-dominated dissipation-less
hydrodynamic regime in harmonic traps with arbitrary anisotropy. We do this  
by giving a systematic treatment of the
linearized hydrodynamic equations based on eqs.(\ref{eq:1.1}) - (\ref{eq:1.6})  
applicable (within our basic assumptions) in the whole temperature range from  
the high temperature domain, where
the Boltzmann limit
$f_{\mp}\simeq \exp(-\beta\left(\frac{\hbar^2k^2}{2m}+V(\vec{x})-\mu\right) )$
applies, to the regime close to the degeneracy temperature for bosons and  
down to nearly vanishing temperature for fermions. We shall discuss a class
of exact solutions of the dissipation-less equations applicable to the whole
temperature-domain covered by the theory, generalizing results obtained  
previously for traps with axial symmetry. It can be shown that
in the high-temperature limit of a classical Boltzmann gas the linearized
hydrodynamic equations in a completely anisotropic trap are integrable and
separable in elliptic coordinates, just as their low temperature counterparts  
\cite{CG2,CG}. However, at lower temperatures where effects of quantum
statistics become important, the integrability and separability are lost,
which manifests itself e.g. in effects of avoided level crossings.

Of special interest in the present paper, besides the common sound modes,
will be the phenomenon of `internal waves', which are characteristic of fluids  
whose equilibrium state is stratified by an external potential. Internal
waves in trapped Fermi gases were mentioned in \cite{BC} but have not yet been  
investigated in detail for trapped quantum-gases. For the discussion of
internal waves in classical contexts like waves in the atmosphere see
\cite{L}.

\section{Linearized hydrodynamic equations and Hilbert-space of their
         solutions}\label{sec:2}
In the present section we write down the five linearized hydrodynamic
equations whose solution is the central theme of this paper. In previous work  
on these equations they were reduced to a set of three wave-equations for the  
velocity field, which we shall also write down for completeness, and some
special solutions of this latter set of equations were given. However, the
appropriate  boundary conditions are hard to formulate for the velocity field  
and therefore it is not clear, so far, which function space is spanned by the  
solutions, and whether a scalar product can be placed on this function space,  
and if so what it is. This question is of particular practical and theoretical  
relevance for the present problem, because in general the solutions have to
be constructed numerically by converting the differential operators to
matrices using a basis and the scalar product in the solution-space. It is
important to choose the correct scalar product  because, as we shall see, the  
problem possesses a dense-lying discrete spectrum of low-lying states, and it  
is a priori far from clear, whether all these states are needed to span the
complete space of solutions, and if not, how the correct states are to be
distinguished. It is our aim here to devote particular attention to this open  
problem and to present an answer. The way to achieve this will be to deviate  
from the previous line of approach by deriving, instead of three coupled
wave-equations for the components of the velocity field, two coupled
wave-equations for the pressure and the density. For these we shall construct  
a scalar product in which the wave-operator is self-adjoint so that its
eigenfunctions form a complete set in a well-defined Hilbert space.

\subsection{Linearized hydrodynamic equations}\label{subsec:2a}
Let us introduce small deviations $\delta\rho$, $\delta P$ of density and
pressure from equilibrium
\begin{eqnarray}
       \rho = \rho_0 \left(\vec{x}\right)+\delta\rho\left(\vec{x},t\right),
       P = P_0\left(\vec{x}\right)+\delta P\left(\vec{x},t\right)
\label{eq:2.1}
\end{eqnarray}
where $\rho_0(\vec{x})$ and $P_0(\vec{x})$ solve the time-independent
hydrodynamic equations with vanishing velocity-field $\vec{u}_0=0$, namely
$\vec{\nabla}P_0(\vec{x})=\rho_0(\vec{x})\vec{f}(\vec{x})$, which defines the  
mechanical equilibrium condition. In principle there are many equilibrium
profiles $\rho_0(\vec{x}), P_0(\vec{x})$ satisfying this requirement. In our  
present context the physically relevant one is the thermodynamic equilibrium  
of maximum local entropy. The entropy-maximum is achieved by the
special profiles $\rho_0(\vec{x}), P_0(\vec{x})$ corresponding to a state
with uniform temperature $T$ and chemical potential $\mu$. Using
eqs.(\ref{eq:1.4}), (\ref{eq:1.5}) $\rho_0(\vec{x})$, $P_0(\vec{x})$
can be written as
\begin{eqnarray}
      \rho_0(\vec{x})
      &&= A_{\mp} m\left(\frac{mk_BT}{2\pi\hbar^2}\right)^{3/2}
      F_{\mp}(\frac{3}{2}, \frac{V(\vec{x})-\mu}{k_BT})
\nonumber\\
      P_0(\vec{x})
      &&= A_{\mp} k_BT \left(\frac{mk_BT}{2\pi\hbar^2}\right)^{3/2}
      F_{\mp}(\frac{5}{2}, \frac{V(\vec{x})-\mu}{k_BT})
\label{eq:2.2}
\end{eqnarray}
where the upper (lower) sign refers to bosons (fermions) and
\begin{eqnarray}
      A_- = 1
           \qquad, \qquad
      A_+ = 2.
\label{eq:2.3}
\end{eqnarray}
Eqs.(\ref{eq:2.2}), but with space-dependent $T(\vec{x})$, $\mu(\vec{x})$ apply
also to the states of local thermodynamic equilibrium, in which the system
always is in the hydrodynamic limit. They can then be taken to define
two of the four fields $P(\vec{x})$, $\rho(\vec{x})$, $T(\vec{x})$,  
$\mu(\vec{x})$ in terms of the other two.
In the fermionic case we need to assume the presence of two equally
populated hyperfine sub-states, the collisions between which can
then ensure the local
thermodynamic equilibrium.
The Bose-Einstein integrals $F_-(s, z)$ and Fermi-Dirac integrals
$F_+(s, z)$ are defined by
\begin{eqnarray}
       F_{\mp}(s, z)=\frac{1}{\Gamma(s)}\int\limits^{\infty}_{0}
         \frac{x^{s-1}}{e^{x+z}\mp1}\, dx
\label{eq:2.4}
\end{eqnarray}
satisfying the familiar recursion relation.
\begin{eqnarray}
       \frac{\partial F_{\mp}(s, z)}{\partial z}
     = -F_{\mp}(s-1, z)
\label{eq:2.4a}
\end{eqnarray}
In the present case $z$, and therefore also $F_{\mp}(s,z)$, is
space-dependent via $z=z(\vec{x})=(V(\vec{x})-\mu)/k_BT$. However,
we shall usually suppress the $z$- and $\vec{x}$-dependence in our
notation for simplicity and just write $F_{\mp}(s)$.

Let us see under which conditions this thermodynamic equilibrium state is
stable against mechanical perturbations. Displacing a volume-element of fluid  
mechanically in the direction of increasing pressure, i.e. in the direction of  
$\vec{f}$, its volume is compressed adiabatically, so that its density is
increased, per unit displacement, by $(\partial\rho_0/\partial  
P_0)_S\vec{\nabla}P_0$, whereas the
density in the ambient equilibrium-gas changes by $\vec{\nabla}\rho_0$ in the  
same displacement. A restoring force   per unit volume in the direction
opposite to the displacement
\begin{equation}
\vec{f}\cdot\left[(\partial\rho_0/\partial P_0)_S\vec{\nabla}P_0
-\vec{\nabla}\rho_0\right]<0
\label{eq:2.4b}
\end{equation}
must result for a mechanically stable state.\footnote{A restoring force does  
not result if the whole fluid-layer on an equi-potential surface is displaced  
in the same way orthogonal to the equi-potential surface; because then no
ambient fluid remains, which could give rise to the buoyancy force
(\ref{eq:2.4b}). Instead a new mechanical equilibrium is reached. This
mechanism gives rise to the zero-frequency modes discussed later.} Using the  
relation $(\partial\rho_0/\partial P_0)_S=3\rho_0/5P_0$, valid for the ideal  
quantum-gases, and eqs.(\ref{eq:2.2})-(\ref{eq:2.4}) we may rewrite
(\ref{eq:2.4b}) as
\begin{equation}
\vec{f}^2\frac{m^2}{(k_BT)^2}\left(\frac{3}{5}
\frac{F_{\mp}^2(\frac{3}{2})}{F_{\mp}(\frac{5}{2})}-F_{\mp}(\frac{1}{2})\right)<0.
\end{equation}
This condition is satisfied for bosons for
$z=(V(\vec{x})-\mu)/k_BT>0$, and for fermions for all positive and negative  
values of $z$.

The hydrodynamic equations linearized in $\vec{u}, \delta\rho, \delta P$ are  
then given by
\begin{eqnarray}
      \rho_0(\vec{x})\partial_t\vec{u}
  &=& -\vec{\nabla}\delta P + \delta\rho\vec{f}(\vec{x})
\label{eq:2.5} \\
      \partial_t\delta P
  &=& -\frac{5}{3}\vec{\nabla}\cdot\left(P_0(\vec{x})\vec{u}\right)
    + \frac{2}{3}\rho_0\left(\vec{x}\right)\vec{f}(\vec{x})\cdot\vec{u}
\label{eq:2.6} \\
      \partial_t\delta\rho
  &=& -\vec{\nabla}\cdot\left(\rho_0(\vec{x})\vec{u}\right)
\label{eq:2.7}
\end{eqnarray}
Eliminating $\delta P$ from eqs.(\ref{eq:2.5}), (\ref{eq:2.6}), then using
the continuity equation (\ref{eq:2.7}) to eliminate also $\delta\rho$, finally  
using that $\vec{\nabla}\rho_0$ is parallel to $\vec{f}$, one obtains
immediately the closed wave-equation for the velocity field
\begin{eqnarray}
      \partial^2_t\vec{u}
    = \frac{5}{3}\frac{P_0(\vec{x})}{\rho_0(\vec{x})}
      \vec{\nabla}\left(\vec{\nabla}\cdot\vec{u}\right)
    + \vec{\nabla}\left(\vec{u}\cdot\vec{f}\right)
    + \frac{2}{3}\vec{f}\left(\vec{\nabla}\cdot\vec{u}\right)
\label{eq:2.8}
\end{eqnarray}
which has been the starting point of previous works \cite{GWS,BC,KPS1,KPS2}. 
As already mentioned we
do not find it the most convenient starting point for our present study as the  
boundary conditions on the velocity field at infinity, the nature of the
function space formed by the solutions, the hermiticity or self-adjoint-ness of  
the wave-operator and hence the nature of the spectrum of eigenvalues all
remain unclear, even if some particular solutions, e.g. those satisfying
$\vec{\nabla}\times\vec{u}=0$, can be constructed. To escape this impasse we
therefore proceed by eliminating the velocity field from eqs.(\ref{eq:2.5} -  
\ref{eq:2.7}) by taking the time-derivative of (\ref{eq:2.7}) and
(\ref{eq:2.6}) and inserting (\ref{eq:2.5}). We obtain after some calculation
\begin{eqnarray}
      \partial^2_t\delta\rho
    = \nabla^2\delta P - \vec{f}(\vec{x})\cdot\vec{\nabla}\delta\rho
                       - (\vec{\nabla}\cdot\vec{f}(\vec{x}))\delta\rho
\label{eq:2.9}
\end{eqnarray}
\begin{eqnarray}
      \partial^2_t\delta P
    = &&\frac{5}{3}\frac{P_0(\vec{x})}{\rho_0(\vec{x})}\nabla^2\delta P
    + \left[
      \frac{5}{3}
      \left(\vec{\nabla}\frac{P_0(\vec{x})}{\rho_0(\vec{x})}\right)
      - \frac{2}{3}\vec{f}(\vec{x})\right]\cdot\vec{\nabla}\delta P
\nonumber
\\    &&-\frac{5}{3}\frac{P_0(\vec{x})}{\rho_0(\vec{x})}\vec{f}(\vec{x})
       \cdot\vec{\nabla}\delta\rho +
       \left[
       \frac{2}{3}\vec{f}^2 - \frac{5}{3}\vec{\nabla}\cdot\left(
       \frac{P_0(\vec{x})}{\rho_0(\vec{x})}\vec{f}(\vec{x})\right)
       \right] \delta\rho
\label{eq:2.10}
\end{eqnarray}
It is convenient to define the function
\begin{eqnarray}
       F_{\mp}(z) = \frac{F_{\mp}(\frac{5}{2},z)}
                          {F_{\mp}(\frac{3}{2},z)}
\label{eq:2.11}
\end{eqnarray}
and its derivative $F^{\prime}_{\mp}(z)$ with respect to its argument
$z$ in terms of which we can write
\begin{eqnarray}
      \frac{P_0(\vec{x})}{\rho_0(\vec{x})}
  &=& \frac{k_BT}{m}F_{\mp}\left((V(\vec{x})-\mu)/k_BT\right)
\label{eq:2.12}\\
      \vec{\nabla}\left(\frac{P_0(\vec{x})}{\rho_0(\vec{x})}\right)
  &=& - \vec{f}(\vec{x})F^{\prime}_{\mp} \left(
      (V(\vec{x})-\mu)/k_BT\right)
\nonumber
\end{eqnarray}
Suppressing in the following the subscript $\mp$  and also the argument
$(V(\vec{x})-\mu)/k_BT$ of $F$ and $F^{\prime}$ for notational simplicity we  
can  rewrite eqs.(\ref{eq:2.9}) - (\ref{eq:2.10}) as
\begin{eqnarray}
      \partial^2_t\delta\rho
    = &&\nabla^2\delta P - \vec{f}\cdot\vec{\nabla}\delta\rho
    - \left(\vec{\nabla}\cdot\vec{f}\right)\delta\rho
\label{eq:2.13}\\
 \partial^2_t\delta P
    = &&\frac{5}{3}\frac{k_BT}{m}F\nabla^2\delta P
    - \left( \frac{5}{3}F^{\prime}
    + \frac{2}{3}\right)\vec{f}\cdot\vec{\nabla}\delta P\nonumber
\\    &&-\frac{5}{3}\frac{k_BT}{m}F\vec{f}\cdot\vec{\nabla}\delta\rho
    +\left[\left(
      \frac{5}{3}F^{\prime}+\frac{2}{3}\right)\vec{f}^2
    -\frac{5}{3}\frac{k_BT}{m}F \left(\vec{\nabla}\cdot \vec{f}\,\right)
     \right]\delta\rho
\label{eq:2.14}
\end{eqnarray}
So far we have gained in simplicity compared to eq.(\ref{eq:2.8}) because we  
have now only two coupled wave-equations instead of three. More important,
however, is the fact that it is clear physically  that the density and
pressure perturbations must go to zero in the limit of large distances from
the center of a confining trap. It should be noted that the same cannot be said
for the velocity field. Indeed, it is clear from (\ref{eq:2.5}) that for
$|\vec{x}|\rightarrow\infty$ where $\rho_0(\vec{x})\rightarrow 0$ the velocity  
field $\vec{u}$ is not necessarily bounded by the hydrodynamic equations.
However, in spite of the improvement of the formulation of the linearized
hydrodynamics we have achieved so far, the self-adjoint-ness of the
wave-operator $\underline{\makebox{\bf H}}$ defined by writing
eqs.(\ref{eq:2.13}), (\ref{eq:2.14}) in the form
\begin{eqnarray}
      \partial^2_t{\delta P\choose\delta\rho}
    = - \underline{\makebox{\bf H}}\cdot{\delta P\choose\delta\rho}
\label{eq:2.15}
\end{eqnarray}
remains to be clarified. Can a scalar product be found in which the operator  
$\underline{\makebox{\bf H}}$ is hermitian? This is the question to which we  
turn next.

\subsection{Scalar product and hermiticity of the wave-operator}
           \label{subsec:2b}
In order to find a useful scalar product on the space of solutions of
eqs.(\ref{eq:2.9}) - (\ref{eq:2.10}) we proceed as follows. First we find a
Lagrangian for eqs.(\ref{eq:2.9}) - (\ref{eq:2.10}), which must be a quadratic  
functional of $\delta\rho(\vec{x},t)$, $\delta P(\vec{x},t)$. It can be found  
by making a general ansatz and comparing the coefficient-functions of the
resulting Euler-Lagrange equations with those in eqs.(\ref{eq:2.9}) -
(\ref{eq:2.10}). From the Lagrangian density we can pass to the associated
`energy density' $\cal{H}$ whose space-integral
\begin{eqnarray}
      E = \int d^3x \cal{H}
\label{eq:2.16}
\end{eqnarray}
must be conserved by time-translation invariance. We shall then define the
scalar product
$\langle\makebox{\bf P}_1|\makebox{\bf P}_2\rangle$
in such a way that the conserved `energy' takes the form
\begin{eqnarray}
      E =
      \langle\makebox{\bf P}|\underline{\makebox{\bf H}}\makebox{\bf P}\rangle
\label{eq:2.17}
\end{eqnarray}
for vectors $\makebox{\bf P}$ satisfying the time-independent wave-equation
$\underline{\makebox{\bf H}}
 \makebox{\bf P}=\omega^2\makebox{\bf P}$.

Thus we begin with the Lagrangian
\begin{eqnarray}
      L = \int d^3x\cal{L}
\label{eq:2.18}
\end{eqnarray}
for whose density we find after some calculation
\begin{eqnarray}
      {\cal L}
    = &&\frac{\beta}{2}\left(\partial_t \delta P\right)^2
    + \alpha\left(\partial_t\delta P\right)\left(\partial_t\delta\rho\right)
    - \frac{5}{3} \frac{k_BT}{m}\frac{F\alpha}{2}(\partial_t\delta\rho)^2
\nonumber
\\  - && \frac{1}{2}\left(\alpha+\frac{5}{3}\frac{k_BT}{m}F\beta\right)
    \left(\vec{\nabla}\delta P\right)^2
    + \vec{E}\cdot\left(\delta\rho\vec{\nabla}\delta P
      - \delta P\vec{\nabla}\delta\rho\right)+I\delta
P\delta\rho+\frac{1}{2}J(\delta\rho)^2
\label{eq:2.19}
\end{eqnarray}
where the coefficients $\alpha, \beta, \vec{E}, I$, and $J$ are defined by
\begin{eqnarray}
      \alpha
   &=& - \frac{K}{(2+5F^{\prime}) F_{\mp}
      \left(\frac{3}{2}, \frac{V-\mu}{k_BT}\right)}
\label{eq:2.20} \\
      \beta
  &=& - \frac{m}{k_BT}\frac{1+F^{\prime}}{F}\alpha
\label{eq:2.21} \\
      \vec{E}
  &=& - \frac{1}{6}\alpha\left(5F^{\prime}+2\right)\vec{f}
\label{eq:2.22}\\
      I &=&
      -\frac{1}{2}\left(\alpha+\frac{5}{3}\frac{k_BT}{m}F\beta\right)
      \vec{\nabla}\cdot\vec{f}+\frac{\beta}{6}\left(5F^{\prime}+2\right)
      \vec{f}^2
\label{eq:2.23} \\
      J &=&
      \frac{\alpha}{3}\left(5F^{\prime}+2\right)\vec f{^2}
\label{eq:2.24}
\end{eqnarray}
The coefficient $K$ in the relation for $\alpha$ is arbitrary, and can be
used for normalization.

\begin{figure}
\centerline{\epsfig{figure=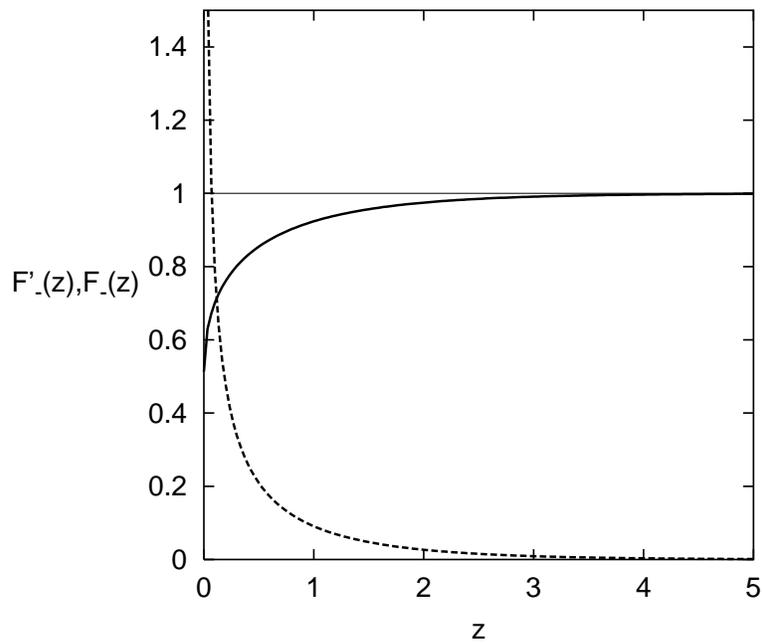,width=10.cm,angle=0}}
\caption{The bosonic function
$F_{-}(z)=F_{-}(\frac{5}{2},z)/F_{-}(\frac{3}{2},z)$ (full
line) and its derivative $F'_{-}(z)$ (broken line) as a function of
$z$.}
\label{fig:1a}
\end{figure}

\begin{figure}
\centerline{\epsfig{figure=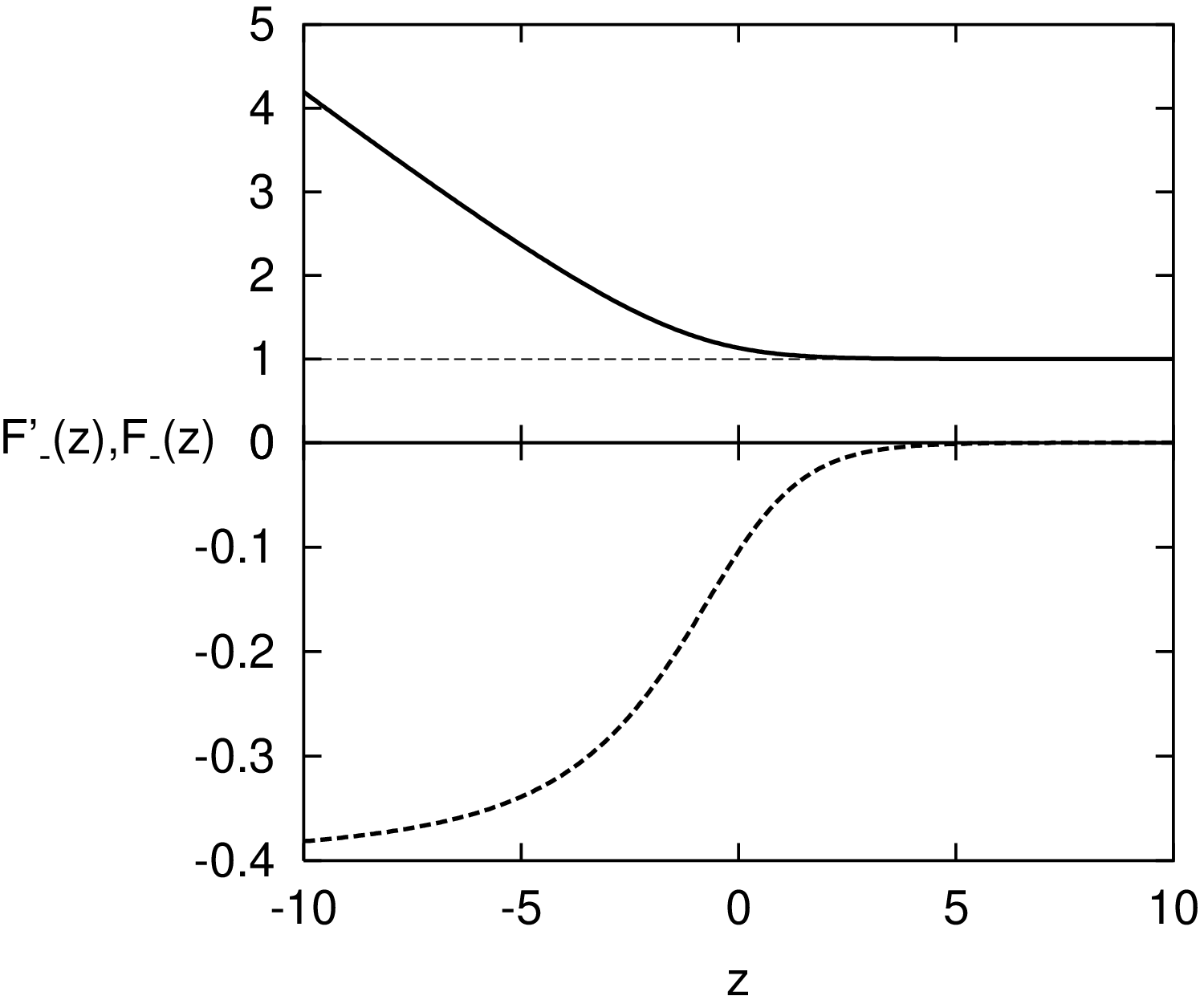,width=10.cm,angle=0}}
\caption{The fermionic function
$F_{+}(z)=F_{+}(\frac{5}{2},z)/F_{+}(\frac{3}{2},z)$ (full
line) and its derivative $F'_{+}(z)$ (broken line) as a function of
$z$.}
\label{fig:1b}
\end{figure}

Next we pass to the associated density ${\cal H}$ defined by the Legendre
transformation
\begin{eqnarray}
      {\cal H} =
      (\partial_t\delta\rho)\frac{\partial{\cal L}}{\partial(\partial_t\delta\rho)}
    + (\partial_t\delta P) \frac{\partial
      {\cal L}}{\partial(\partial_t\delta P)} - {\cal L},
\label{eq:2.25}
\end{eqnarray}
with the conserved quantity (\ref{eq:2.16}).
It is now useful to employ the vector-notation already defined in
eq.(\ref{eq:2.15}) by defining
\begin{eqnarray}
      \makebox{\bf P}(\vec{x}, t) =
      {\delta P (\vec{x}, t)
      \choose \delta\rho(\vec{x}, t)}.
\label{eq:2.27}
\end{eqnarray}
With the harmonic time-dependence
\begin{eqnarray}
      \makebox{\bf P}(\vec{x}, t) =
      \makebox{\bf P}(\vec{x})\cos(\omega t+\varphi)
\label{eq:2.28}
\end{eqnarray}
the conserved quantity $E$ can be written as
\begin{eqnarray}
      E = &&\omega^2\sin^2(\omega t+\varphi)\int d^3x
      \left[\frac{\beta}{2}(\delta P)^2 +\alpha\delta P\delta\rho
    - \frac{5}{3}\frac{k_BT}{m}\frac{F\alpha}{2}(\delta\rho)^2\right]
\nonumber \\
    &&+\cos^2(\omega t+\varphi)\int d^3x \left[\frac{1}{2}\left(\alpha
    +\frac{5}{3}\frac{k_BT}{m}F\beta\right)\left(\vec{\nabla}\delta
     P\right)^2\label{eq:2.29}\right.\\
     &&\left.~~~~~~~~~~~~~~~~~~~~~~~~~~~
-\vec{E}\cdot\left(\delta\rho\vec{\nabla}\delta P
    -\delta P\vec{\nabla}\delta\rho\right)-I\delta P\delta\rho
    -\frac{1}{2}J(\delta\rho)^2\right]\nonumber
\end{eqnarray}
In order to meet our goal (\ref{eq:2.17}) we should define the scalar product  
$\langle\makebox{\bf P}_1 |\makebox{\bf P}_2\rangle$ in such a way that the
coefficients of $\sin^2$ and $\cos^2$ in (\ref{eq:2.29}) become both equal to  
$\omega^2\langle\makebox{\bf P} |\makebox{\bf P}\rangle$. Therefore, we can
conclude from the coefficient of $\sin^2$ that the norm becomes
\begin{eqnarray}
      \langle \makebox{\bf P} |\makebox{\bf P}\rangle
    = \int d^3x\left[
      \frac{\beta}{2}\delta P^2 +\alpha\delta P\delta\rho
    - \frac{5}{6}\frac{k_BT}{m}F\alpha(\delta\rho)^2\right]
\label{eq:2.30}
\end{eqnarray}
Using the relations (\ref{eq:2.20}), (\ref{eq:2.21}) for $\alpha$ and $\beta$  
it can be checked that the norm is positive, as required, if $\alpha < 0$,
(which can always be achieved by the choice of the constant $K$), and
$\beta > 0$,
which requires the inequality $1+F^{\prime}>0$, and
$-\frac{5}{3}\frac{k_BT}{m}F\alpha\beta -\alpha^2 >0$ which in turn
requires the
stronger inequality
\begin{eqnarray}
      5F^{\prime}((V(\vec{x})-\mu)/ k_B T) + 2 > 0
\label{eq:2.31}
\end{eqnarray}
Using the definition (\ref{eq:2.11}) of $F=F_{\mp}$ it is easy to check that  
(\ref{eq:2.31}) is equivalent to the stability condition (\ref{eq:2.4b}) of
the thermodynamic equilibrium state. The functions $F(z),
F^{\prime}(z)$
are plotted in figs.\ref{fig:1a} for bosons and \ref{fig:1b} for fermions.
  From eq.(\ref{eq:2.30}) we deduce that the scalar product on the complexified  
space of solutions can be defined as
\begin{eqnarray}
      \langle \makebox{\bf P}_1 |\makebox{\bf P}_2\rangle
    = \int d^3x \left(\delta P^*_1, \delta\rho^*_1\right)
      \underline{{\bf S}}{\delta P_2 \choose\delta\rho_2}
\label{eq:2.32}
\end{eqnarray}
with the matrix
\begin{eqnarray}
      \underline{{\bf S}} =
      \left({\frac{\beta}{2}\atop\frac{\alpha}{2}}
            {\frac{\alpha}{2}\atop-\frac{5}{3}}
            {\atop\frac{k_BT}{m}
            \frac{F\alpha}{2}}\right)
\label{eq:2.33}
\end{eqnarray}
The coefficient of $\cos^2$ in eq.(\ref{eq:2.29}) can now be checked to be of  
the form
\begin{eqnarray}
      \langle
      \makebox{\bf P} | \underline{{\bf H}}\makebox{\bf P}\rangle
    = \omega^2\langle\makebox{\bf P} | \makebox{\bf P}\rangle
\label{eq:2.34}
\end{eqnarray}
for vectors $|\makebox{\bf P}\rangle$ satisfying
    $\underline{{\bf H}} |\makebox{\bf P}\rangle = \omega^2 |
     \makebox{\bf P}\rangle$.
Indeed, restricting to a real space, for simplicity, because $\underline{{\bf  
H}}$ is real, we find by direct evaluation
\begin{eqnarray}
      \langle
      \makebox{\bf P}|\underline{{\bf H}}\makebox{\bf P}\rangle &=&
      \int d^3x(\delta P,\delta\rho)\underline{{\bf S}}\cdot
      \underline{{\bf H}}{\delta P \choose \delta\rho}
\nonumber\\
&=& \int d^3x \left[
      \frac{1}{2}\left(\alpha+\frac{5}{3}\frac{k_BT}{m}F\beta\right)
      \left(\vec{\nabla}\delta P\right)^2
    - \vec{E}\cdot\left(\delta\rho\vec{\nabla}\delta P-\delta P\vec{\nabla}
      \delta\rho\right)
    - I\delta P\delta\rho-\frac{J}{2}(\delta\rho)^2\right] \nonumber\\
\label{eq:2.36}
\end{eqnarray}
We can furthermore show by direct calculation that the wave-operator is
hermitian in the scalar product (\ref{eq:2.32}), if suitable boundary
conditions are imposed. We find after straightforward, but lengthy calculation
\begin{eqnarray}
      \langle
      \makebox{\bf P}_1 |\underline{{\bf H}}\makebox{\bf P}_2\rangle
    - \langle\makebox{\bf P}_2 |\underline{{\bf H}}\makebox{\bf P}_1\rangle
    &&= \int d^3x
      \left(\makebox{\bf P}_1\underline{{\bf S}}\cdot
      \underline{{\bf H}}\makebox{\bf P}_2
    - \makebox{\bf P}_2\underline{{\bf S}} \cdot
      \underline{{\bf H}}\makebox{\bf P}_1\right)
\nonumber   \\
  &&= \frac{1}{2}\int d^3x\vec{\nabla}\cdot
      \left\{
      (\alpha+\frac{5k_BT}{3m}F\beta)\right.\label{eq:2.38}\\
      &&\left.~~~~~~~~~~~~~~~\left[(
      \delta P_1\vec{\nabla}\delta P_2-\delta P_2\vec{\nabla}\delta P_1
      )
    - \vec{f}\left(\delta P_1\delta\rho_2-\delta P_2\delta\rho_1
      \right)\right]\right\}\nonumber
\end{eqnarray}
Thus, we must impose boundary conditions at infinity in such a way that the
surface-integral we obtain from (\ref{eq:2.38}) by the application of Gauss'  
law vanishes. Since the coefficient functions $\alpha$ and $\beta$ grow for
$|\vec{x}|\rightarrow\infty$ like
$\exp\left(\frac{V(\vec{x})-\mu}{k_BT}\right)$, the fluctuations $\delta
P(\vec{x})$ and $\delta\rho(\vec{x})$ for all solutions must vanish
sufficiently rapidly for $|\vec{x}|\rightarrow\infty$.

Finally, let us transform the scalar product (\ref{eq:2.32}) to the more
symmetrical form
\begin{eqnarray}
      \langle\makebox{\bf P}_1|\makebox{\bf P}_2\rangle
    = \int d^3x\left(u^*_1u_2 + v^*_1v_2\right)
\label{eq:2.39}
\end{eqnarray}
by the linear transformation
\begin{eqnarray}
      \makebox{\bf P}
    = {\delta P\choose\delta\rho} = \underline{{\bf M}} {u\choose v}
\label{eq:2.40}
\end{eqnarray}
with a matrix $\underline{{\bf M}}$ in lower triangular form
\begin{eqnarray}
      \underline{{\bf M}} = \left({a \atop b} {0 \atop c}\right)
\label{eq:2.41}
\end{eqnarray}
diagonalizing and normalizing the kinetic term in the Lagrangian $\cal{L}$
(and hence also in the `energy-density' $\cal{H}$). The latter requirement
yields
\begin{eqnarray}
    a &=&
         \sqrt{\frac{5k_BT}{m}}\sqrt{\frac{-F}{\alpha(5F^{\prime}+2)}}
\nonumber \\
    b &=&
         \sqrt{\frac{9m}{5k_BT}}\frac{1}{\sqrt{-\alpha F(5F^{\prime}+2)}}
\label{eq:2.42} \\
    c &=&
          \sqrt{\frac{3m}{5k_BT}}\frac{1}{\sqrt{-\alpha F}}
\nonumber
\end{eqnarray}
In the new variables $u, v$ the wave-equation now reads
\begin{eqnarray}
    - \frac{\partial^2}{\partial t^2}
      {u\choose v} = \underline{{\bf \tilde{H}}}{u\choose v}
\label{eq:2.43}
\end{eqnarray}
with the manifestly hermitian wave-operator
\begin{eqnarray}
      \underline{{\bf \tilde{H}}} = \left(
      \begin{array}{ccc}
       {-\frac{5}{3}\frac{k_BT\vec{\nabla}\cdot F\vec{\nabla}}{m}
      +\frac{\left(\vec{\nabla}\cdot\vec{f}\right)}{6}
      +\frac{m}{60k_BT}\frac{\vec{f}^2}{F},}
      &&
      {\frac{\vec{\nabla}\cdot\sqrt{5F^{\prime}+2}\vec{f}}{\sqrt{3}}
      +\frac{\sqrt{3}m}{30k_BT}
      \frac{\sqrt{5F^{\prime}+2}\vec{f}^2}{F},}\\
      &&  \\
      {-\frac{\sqrt{5F^{\prime}+2}\vec{f}\cdot\vec{\nabla}}{\sqrt{3}}
      +\frac{\sqrt{3}m}{30k_BT}\frac{\sqrt{5F^{\prime}+2}\vec{f}^2}{F},}

      &&
      {\frac{m}{5k_BT}\frac{(5F^{\prime}+2)\vec{f}^2}{F}}
      \end{array}\right)
\label{eq:2.44}
\end{eqnarray}

\subsection{Zero-frequency modes and isothermal modes}\label{subsec:2c}
For arbitrary temperature $T$ and trap-potential $V(\vec{x})$
eqs.(\ref{eq:2.9}), (\ref{eq:2.10}) possess a class of exact time-independent  
solutions, which depend on an arbitrary function $G(V(\vec{x}))$, and its
derivative $G^{\prime}=dG/dV$, namely
\begin{eqnarray}
      \delta\rho(\vec{x})
   &=& -\varepsilon G^{\prime}(V(\vec{x}))
\nonumber \\
      \delta P(\vec{x})
   &=& \frac{\varepsilon}{m} G(V(\vec{x})).
\label{eq:A}
\end{eqnarray}
$\varepsilon$ is a parameter which is sufficiently small to make the
linearized theory consistent. The norm (\ref{eq:2.30}) of these solutions
$|\makebox{\bf P}_0\rangle$ is
\begin{eqnarray}
      \langle\makebox{\bf P}_0|\makebox{\bf P}_0\rangle
    = \varepsilon^2\int d^3x\frac{|\alpha|}{2m} \left[
      \frac{1+F^{\prime}}{k_BTF}G^2+2GG^{\prime}+\frac{5}{3}
      k_BTF G^{\prime 2}\right].
\label{eq:B}
\end{eqnarray}
Since all functions under the integral depend on $\vec{x}$ only via
$V(\vec{x})$ the integration $\int d^3x...$ can be replaced by $const\int
dV\sqrt{V}...$, if V is a homogeneous function of second order of $\vec{x}$,
e.g. a parabolic potential. The scalar product exists and is positive under
the condition (\ref{eq:2.31}), if $G(V)$ vanishes sufficiently rapidly, e.g.  
like $\exp(-V/k_BT)$, for $V\rightarrow\infty$. Then these solutions belong to  
the Hilbert-space and have to be considered. Physically, they appear because  
of the coexistence of a continuum of mechanical equilibrium states and the
thermodynamic equilibrium. As we shall see in the following section these
states are not isolated from all the other states but occur, for any local
wave-number, as the end-point of a spectral branch of states if the local
wave-number is turned in the direction of $\vec{f}=-\frac{1}{m}\vec{\nabla}V$.

There are some further exact solutions of the wave-equations (\ref{eq:2.13},
\ref{eq:2.14}) which hold for all temperatures in the fermionic -and all
temperature $T>T_c$ in the bosonic case. They are obtained by extending the  
ansatz (\ref{eq:A}) for the zero-frequency modes according to
\begin{eqnarray}
      {\delta P(\vec{x},t)\choose\delta\rho(\vec{x},t)}
    = \epsilon x^\alpha y^\beta z^\gamma{\frac{1}{m}G(V(\vec{x}))\choose  
-G'(V(\vec{x}))}e^{-i\omega t}
\label{eq:C}
\end{eqnarray}
with $\alpha$, $\beta$, $\gamma = 0$ or $1$. Inserting this ansatz in  
eq.(\ref{eq:2.13}), and using the property that for $\alpha=\beta=\gamma=0$  
eq.(\ref{eq:C}) is a zero-frequency mode, we find after a simple calculation
\begin{equation}
(\omega^2-\alpha\omega_1^2-\beta\omega_2^2-\gamma\omega_3^2)G'(V)=0.
\label{eq:D}
\end{equation}
Next we insert the ansatz also in eq.(\ref{eq:2.14}) and obtain by a similar  
calculation
\begin{equation}
\left(\omega^2+(\alpha\omega_1^2+\beta\omega_2^2+\gamma\omega_3^2)(\frac{5}{3}F'+\frac{2}{3})\right)G(V)+\frac{5}{3}k_BT(\alpha\omega_1^2+\beta\omega_2^2+\gamma\omega_3^2)FG'(V)=0.
\label{eq:E}
\end{equation}
Eq.(\ref{eq:D}) determines the mode-frequencies as
\begin{equation}
\omega_{\alpha\beta\gamma}=\sqrt{\alpha\omega_1^2+\beta\omega_2^2
+\gamma\omega_3^2},
\label{eq:F}
\end{equation}
while (\ref{eq:E}), for $\alpha, \beta, \gamma$ not all equal to zero, fixes  
the yet undetermined function $G(V)$ in the ansatz (\ref{eq:C}) as
\begin{equation}
G(V)=const F(\frac{3}{2},\frac{V(\vec{x})-\mu}{k_BT}).
\label{eq:G}
\end{equation}
It follows with (\ref{eq:C}) that $\delta P(\vec{x},t)$ and  
$\delta\rho(\vec{x},t)$ for these modes are related by
\begin{equation}
\delta  
P(\vec{x},t)=\frac{k_BT}{m}\frac{F(\frac{3}{2},\frac{V(\vec{x})-\mu}{k_BT})}{F(\frac{1}{2},\frac{V(\vec{x})-\mu}{k_BT})}\delta\rho(\vec{x},t),
\end{equation}
which is the relation between changes of pressure and density implied by the
local thermodynamic equilibrium (\ref{eq:2.2}) if the temperature is kept  
constant. These isothermal modes were already found in \cite{GWS} for
the special case of isotropic and axially symmetric parabolic trap-potentials.

The modes (\ref{eq:F}) contain as special cases the three Kohn-modes
$\omega_{100}=\omega_1, \omega_{010}=\omega_2, \omega_{001}=\omega_3$,  
corresponding to oscillations of the center of mass of the trapped gas.
It is interesting to note that collisionless Kohn-modes of the form
\begin{equation}
\delta\rho(\vec{x},t)=\epsilon\frac{\partial}{\partial  
x_i}\rho_0(\vec{x})e^{-i\omega t}
\end{equation}
with the same frequencies $\omega_i$ also exist. It therefore follows from
the phenomenological formula (\ref{int}) that these modes are not damped by  
the relaxation mechanisms present in the system, in agreement with the
general statement made by the Kohn-theorem.

For fermions the result (\ref{eq:F}) and (\ref{eq:G}), (\ref{eq:C}) for the  
frequencies and mode-functions apply to all temperatures and can therefore
also be extrapolated to $T\rightarrow 0$. Indeed, for $T\rightarrow 0$,
 modes with the frequencies (\ref{eq:F}) where found in \cite{CG}.
In order to compare also the mode-functions we use the Bethe-Sommerfeld  
expansion to evaluate the Fermi-integrals asymptotically $F_+(s,z)\sim  
(-z)^s/s!$ for $z\rightarrow -\infty$ and find for $T\rightarrow 0$
\begin{equation}
\delta P =\frac{2}{3}\frac{\mu-V(\vec{x})}{m}\delta\rho
\sim x^\alpha y^\beta z^\gamma\left(\mu-V(\vec{x})\right)^{3/2},
\end{equation}
in agreement with \cite{CG}.

Finally, another set of solutions for arbitrary temperature is found by  
generalizing an ansatz of \cite{GWS} for a trap without axial symmetry by  
putting
\begin{equation}
u_i(\vec{x},t) =A_i x_i e^{-i\omega t}
\end{equation}
with three constants $A_i$.
Inserting directly into eqs.(\ref{eq:2.6},\ref{eq:2.7}) we obtain the density  
and pressure modes
\begin{eqnarray}
\delta \rho(\vec{x},t)&&=-{im \over \omega}
  \left({mk_B T \over 2\pi \hbar^2} \right)^{3/2} 
    \left[ F\left({3 \over 2}\right)\sum_{i=1}^3 A_i
-F\left({1 \over 2}\right) {m \over k_B T}
\sum_{i=1}^3 A_i \omega_i^2 x_i^2\right]e^{-i\omega t}.\label{deltaT1}\\
\delta P(\vec{x},t)&&={ik_B T \over \omega} 
\left({mk_B T \over 2\pi \hbar^2} \right)^{3/2}
\left[-{ 5 \over 3}
F\left({5 \over 2}\right)\sum_{i=1}^3 A_i
-F\left({3 \over 2}\right) {m \over k_B T}
\sum_{i=1}^3 A_i \omega_i^2 x_i^2\right]e^{-i\omega t}.
\label{deltaT2}
\end{eqnarray}
A comparison of (\ref{deltaT1})-(\ref{deltaT2}) with the local equilibrium relations (\ref{eq:1.4})-(\ref{eq:1.5}) reveals that the temperature oscillates in these modes with a spatially constant amplitude proportional to $\sum_{i=1}^3 A_i$.
Finally, using the results (\ref{deltaT1}),(\ref{deltaT2}) in the equation (\ref{eq:2.5}) for momentum
conservation
we arrive at the eigenvalue problem:
\begin{equation}
\omega^2 A_i=2 A_i \omega_i^2 + {2 \over 3}
\omega_i^2 \sum_{j=1}^3 A_j \qquad i=1,2,3.
\label{eq:cubic}
\end{equation}
The eigenvector $\vec{A}$ and the eigenvalue $\omega^2$ are
clearly temperature independent and follow from the cubic
secular equation
\begin{eqnarray}
      (\omega^2)^3 -\frac{8}{3}
                   \left(\omega^2_x+\omega^2_y+\omega^2_z\right)
      (\omega^2)^2 + \frac{20}{3}
                   \left(\omega^2_x\omega^2_y+\omega^2_x\omega^2_z
                   +\omega^2_y\omega^2_z\right)
          \omega^2 - 16\omega^2_x\omega^2_y\omega^2_z = 0.
\label{eq:e}
\end{eqnarray}
In the special case of an axially symmetric trap the cubic equation can be  
reduced to a quadratic one and a result first obtained in \cite{GWS}
is recovered.

\section{Short wave-length solutions}\label{sec:3}
The two coupled wave-equations derived in the previous section in various
forms are difficult to solve for arbitrary temperature in a system which is
made spatially inhomogeneous by an external potential $V(\vec{x})\neq 0$. An  
exception, however, are waves of wave-lengths, which are short on the spatial  
scale on which $V(\vec{x})$ and hence also $P_0(\vec{x}), \rho_0(\vec{x})$
vary. Such waves, in the representation with $u(\vec{x},t), v(\vec{x},t)$,
can be written as
\begin{eqnarray}
      {u(\vec{x},t)\choose v(\vec{x},t)    }
    = e^{-i\omega t}
      {a_0(\vec{x})\choose b_0(\vec{x})    }e^{iS(\vec{x},t)}
\label{eq:3.1}
\end{eqnarray}
The eikonal $S(\vec{x},t)$ defines the local wave-vector by the relation
\begin{eqnarray}
      \vec{k}(\vec{x},t) = \vec{\nabla} S(\vec{x},t)
\label{eq:3.2}
\end{eqnarray}
The amplitudes $a_0, b_0$ and also $\vec{k}$ vary slowly in space, on the
same scale as $V(\vec{x}), P_0(\vec{x}), \rho_0(\vec{x})$. The frequency
$\omega$ is independent of $\vec{x},t$. Inserting the ansatz in the equation  
(\ref{eq:2.43}) and neglecting derivatives of $a_0, b_0$ and $\vec{k}$, and
assuming also that $\omega^2\gg|\vec{\nabla}\cdot\vec{f}|$, we obtain the
secular equation as the vanishing of the determinant
\begin{eqnarray}
      \left|
      \begin{array}{ccc}
      {\frac{5}{3}\frac{k_BT}{m}F\vec{k}^2
    + \frac{1}{60}\frac{m}{k_BT}\frac{\vec{f}^2}{F}-\omega^2
      }
      &&
      {\frac{i}{\sqrt{3}}\sqrt{5F^{\prime}+2}(\vec{k}\cdot\vec{f})
    + \frac{\sqrt{3}}{30}\frac{\sqrt{5F^{\prime}+2}}{F}\frac{m}{k_BT}
      \vec{f}^2
      }\\
      &&  \\
      {- \frac{i}{\sqrt{3}}\sqrt{5F^{\prime}+2}(\vec{k}\cdot\vec{f})
    + \frac{\sqrt{3}}{30}\frac{\sqrt{5F^{\prime}+2}}{F}\frac{m}{k_BT}
      \vec{f}^2
      }
      &&
      {\frac{5F^{\prime}+2}{5F}\frac{m}{k_BT}\vec{f}^2-\omega^2
      }
      \end{array}\right|
      &=  0
\label{eq:3.3}
\end{eqnarray}
We note that terms with $\vec{f}^2$ and $\vec{k}\cdot\vec{f}$ are essential
to keep in this approximation together with the $k^2$-terms, because
$|\vec{f}|$ grows at large distances from the trap-center, at least for
parabolic traps, and provides the physically crucial confining mechanism. On  
the other hand, $a_0(\vec{x}), b_0(\vec{x})$, and $\vec{\nabla}\cdot\vec{f}$  
do not grow in a similar way and are therefore consistently negligible.

 From eq.(\ref{eq:3.3}) we deduce the local dispersion-law for waves of
short-wavelength $2\pi/k$
\begin{eqnarray}
      \omega^2 =&&
      \omega^2_{\pm}(\vec{k},\vec{x}) = \frac{1}{2}
      \left[
      \frac{5}{3} \frac{k_BT}{m}F k^2 + \frac{m}{k_BT}
      \left(\frac{1}{60F}+\frac{5F^{\prime}+2}{5}\right)\vec{f}^2\right]
\nonumber \\
    &&\pm\sqrt{\frac{1}{4}
      \left[
      \frac{5}{3} \frac{k_BT}{m}F k^2 + \frac{m}{k_BT}
      \left(\frac{1}{60F}+\frac{5F^{\prime}+2}{5}\right)\vec{f}^2\right]^2
    - \frac{1}{3}\left(5F^{\prime}+2\right)
      \left(\vec{f}\times\vec{k}\right)^2}
\label{eq:3.4}
\end{eqnarray}
In the same level of approximation the pressure and density oscillations are  
related by
\begin{eqnarray}
      \delta P(\vec{k},x) = \frac{1}{k^2}
      \left[\omega^2 - i\vec{f}\cdot\vec{k}\right]
      \delta\rho(\vec{k},x)
\label{eq:3.5}
\end{eqnarray}
as follows from eq.(\ref{eq:2.9}).

The dispersion law (\ref{eq:3.4}) contains a lot of physics and will be
discussed now. First we note that for $\vec{f}\ne 0$ there are two branches of  
the dispersion law, one of high frequency  and another one of lower
frequency, which are both physical. Thus, there are two different types of
waves in these systems. Both branches correspond to frequencies $\omega^2\ge  
0$ for all $\vec{k}$, i.e. to stable oscillation waves.
Another simple observation is that the local dispersion-relation is
anisotropic and depends on the angle between $\vec{f}$ and $\vec{k}$. The
physical nature of the two branches is most easily seen by assuming that the  
angle between $\vec{f}$ and $\vec{k}$ is sufficiently small to permit the
expansion of the square-root in (\ref{eq:3.4}) in the second term of its
radicand. We obtain then to lowest  non-vanishing order
\begin{eqnarray}
       \omega^2_+ =\frac{5}{3}\frac{k_BT}{m} F k^2
     + \frac{m}{k_BT}
       \left(\frac{1}{60F}+\frac{5F^{\prime}+2}{5}\right)\vec{f}^2
\label{eq:3.6}
\end{eqnarray}
\begin{eqnarray}
       \omega^2_- = \frac{5F^{\prime}+2}{3}
       \frac{(\vec{f}\times\vec{k})^2}
            {\frac{5}{3}\frac{k_BT}{m}Fk^2+\frac{m}{k_BT}
            \left(\frac{1}{60F}+\frac{5F^{\prime} +2}{5}\right)
            \vec{f}^2}
\label{eq:3.7}
\end{eqnarray}
 The high-frequency branch is easily recognized in this limit as an adiabatic  
sound mode, in particular if the identity for the adiabatic sound velocity
\begin{eqnarray}
      c^2_s = \left.\frac{\partial P_0}{\partial\rho_0}\right|_s
            = \frac{5}{3}\frac{P_0}{\rho_0}
            = \frac{5}{3}\frac{k_BT}{m}F
\label{eq:3.8}
\end{eqnarray}
is used, which is valid for the ideal quantum gases, with the definition of
$F=F_{\mp}$ by eqs.(\ref{eq:2.11}), (\ref{eq:2.12}). The low-frequency branch  
reaches its lowest frequency $\omega_-=0$, in the present approximation, for  
all waves traveling locally in the direction parallel to the force of the
trap $\vec{f}$, so that $\vec{f}\times\vec{k}=0$. The existence of such
zero-frequency modes has already been discussed in the preceding section.
Looking at exact solutions at high-temperature in the next section we shall
see them appear again.

The low-frequency branch for given $|\vec{k}|$ achieves its highest frequency  
if the wave propagates locally in directions orthogonal to $\vec{f}$. The
maximum frequency for modes orthogonal to $\vec{f}$ is then reached for short  
wave-lengths
\begin{eqnarray}
      c^2_s k^2 \gg \frac{m}{k_BT}
          \left(\frac{1}{60F}+F^{\prime}+\frac{2}{5}\right)\vec{f}^2
\label{eq:3.9}
\end{eqnarray}
and given by
\begin{eqnarray}
      \omega^{max}_-=\sqrt{\frac{5F^{\prime}+2}{3}}\frac{|\vec{f}|}{c_s}.
\label{eq:3.10}
\end{eqnarray}
Waves with the properties of the low-frequency branch solutions found here
are typical for media which are stratified by an external force and are called  
`internal waves'. One of their surprising and counter-intuitive properties is  
that in regions where (\ref{eq:3.9}) applies the group-velocity
$\vec{\nabla}_k\omega_-(\vec{k})$ is orthogonal to the wave-vector $\vec{k}$.  
For a text-book discussion of such waves see \cite{L}. Indeed, the  
dispersion-relations (\ref{eq:3.6}),(\ref{eq:3.7}) can e.g. be directly  
compared with eqs.(53),(54) given there.

\section{Solution in the high-temperature region}\label{sec:4}
\subsection{Hilbert-space of polynomial solutions}\label{subsec:4a}

The high-temperature regime is defined by
\begin{eqnarray}
      T \gg T_{deg}
\label{eq:4.1}
\end{eqnarray}
where $T_{deg}$ is the degeneracy temperature at which the de Broglie
wave-length becomes of the order of the mean particle distance. In this
regime we can approximate
\begin{eqnarray}
      f_{\mp}(\vec{p},\vec{x})
    = e^{-(\frac{p^2}{2m}+V(\vec{x})-\mu) /k_BT}
\label{eq:4.2}
\end{eqnarray}
and $\rho_0(\vec{x})=\rho_0(0)e^{-V(\vec{x})/k_BT}$,
    $P_0(\vec{x})=\frac{k_BT}{m}\rho_0(\vec{x})$. Is is useful to introduce
    new dimensionless variables $P_1(\vec{x}), \rho_1(\vec{x})$ via the definitions
\begin{eqnarray}
      \delta P({\vec{x},t})
   &=& \frac{k_BT}{m}\rho_0(\vec{x}) P_1(\vec{x},t)
\nonumber \\
      \delta\rho({\vec{x},t})
   &=& \rho_0(\vec{x})\rho_1(\vec{x},t)
\label{eq:4.3}
\end{eqnarray}
and to rewrite the coupled wave-equations for $\delta P$, $\delta\rho$ as
coupled equations for $P_1$ and
\begin{eqnarray}
      P_1-\rho_1 = T_1 = \frac{\delta T}{T}
\label{eq:4.4}
\end{eqnarray}
where $\delta T (\vec{x},t)$ is the deviation of the temperature from
equilibrium. Separating the time-dependence $ e^{-i\omega t}$ we
arrive at
\begin{eqnarray}
      \omega^2
           {P_1 \choose P_1-\rho_1}
    = \left(
      \begin{array}{ccc}
      {-\frac{k_BT}{m}\frac{5}{3}\nabla^2-\vec{f}\cdot\vec{\nabla},
      } &&
      {-\frac{5}{3}\left[
      \left(\vec{\nabla}\cdot\vec{f}\right)+\vec{f}\cdot\vec{\nabla}\right]
      -\frac{m}{k_BT}\vec{f}^2
      }\\
      &&\\
      {-\frac{k_BT}{m}\frac{2}{3}\nabla^2,
      } &&
      {-\frac{2}{3}\left[
      \left(\vec{\nabla}\cdot \vec{f}\,\right)+\vec{f}\cdot\vec{\nabla}\right]
      }\end{array}\right)
      {P_1 \choose P_1-\rho_1}
\label{eq:4.5}
\end{eqnarray}
We specialize these equations to a harmonic potential
\begin{eqnarray}
      V(\vec{x}) = \frac{m}{2}
       \left(\omega_x^2 x^2 + \omega_y^2y^2 + \omega_z^2z^2\right)
\label{eq:4.6}
\end{eqnarray}
It is then clear, via term by term inspection, that there are polynomial
solutions of eq.(\ref{eq:4.5}), in which $P_1$ is  polynomial in the
Cartesian components of $\vec{x}$ of total order $n$ and $P_1-\rho_1$
is a corresponding polynomial of total order $n-2$. This is so because
then each term on the right-hand side of (\ref{eq:4.5}) either decreases the  
total order
of the polynomial on the left-hand side by 2 via the operation of $\nabla^2$
or keeps the same order of the polynomial. It follows that we can pick
freely as the highest total power an integer $n$ for $P_1$ and determine the  
eigenvalues
$\omega^2_n$ by comparing the coefficients of all terms with this highest
total power, imposing the condition for nontrivial solvability. It is clear
that we get polynomials of arbitrary total order in this way. Moreover
these solutions lie in the Hilbert-space because via eq.(\ref{eq:4.3})
$\delta P$ and $\delta\rho$ fall off sufficiently rapidly for
$|\vec{x}|\rightarrow\infty$, if $P_1$ and $\rho_1$ are polynomials in
the Cartesian components of $\vec{x}$.

To see this explicitly we specialize the scalar product (\ref{eq:2.32}) for the
present high-temperature case by replacing
$F(3/2,(V(\vec{x})-\mu)/k_B T)=\exp(-(V(\vec{x})-\mu)/k_BT)$,
$ F_{\mp}=1$, $F_{\mp}^{\prime}=0$ which gives
$\alpha=-(K/2)\exp((V(\vec{x})-\mu)/k_BT), \beta=-(m/k_BT)\alpha$, and the
scalar product
\begin{eqnarray}
      \langle\makebox{\bf P}|\tilde{\makebox{\bf P}}\rangle
    = \frac{K}{4}\frac{k_BT}{m}\left(\rho_0(0)\right)^2
      e^{-{\mu \over k_B T}}
      \int d^3x e^{-\frac{V(\vec{x})}{k_BT}}\left[
      P_1\tilde{P}_1-P_1\tilde{\rho}_1-\rho_1\tilde{P}_1
      +\frac{5}{3}\rho_1\tilde{\rho}_1\right]
\label{eq:4.6a}
\end{eqnarray}
It is also clear now, that for harmonic potentials $V(\vec{x})$ the
polynomial solutions for $P_1$ and $\rho_1$ of all orders are complete in a
space with this scalar product, a fact which is very familiar from the quantum  
mechanics of the harmonic oscillator.

To be  specific let us consider a polynomial of $x,y,z$ for $P_1$
of total order $n$. It has then terms of highest total order of the form

\begin{eqnarray}
      P_1 = \sum_{n_1,n_2,n_3}A_{n_1n_2n_3}
              x^{n_1}  y^{n_2}  z^{n_3} + \makebox{lower order}
\label{eq:4.6b}
\end{eqnarray}
with $n_1+n_2+n_3=n$, and ${n+2\choose 2}=(n+2)(n+1)/2$ different
coefficients$A_{n_1n_2n_3}$. For the same mode $P_1-\rho_1$ must be a
polynomial of total order $n-2$ with ${n\choose 2}=n(n-1)/2$ terms $x^{n_1}
y^{n_2} z^{n_3}$ of highest total order $n_1+n_2+n_3=n-2$ with coefficients
$B_{n_1 n_2 n_3}$.

The solvability-condition for the linear homogeneous equations connecting all  
these coefficients then gives a secular equation for the eigenvalues
$\omega^2_n$ of order
\begin{eqnarray}
      {n+2\choose 2}+{n\choose 2} = n^2+n+1
\label{eq:4.6c}
\end{eqnarray}
with just as many solutions. $(n+1)(n+2)/2$ of these modes can be considered  
as modes of sound waves, modified by the external potential, while the
remaining $n(n-1)/2$ can be considered as modes of internal waves, modified by  
their coupling to sound waves. The second kind of modes therefore exists only  
for $n\ge2$.

How many linearly independent zero-frequency modes appear at a given total
order $n$? If $n$ is odd, the polynomials for $P_1$ and $\rho_1$ have odd
parity and cannot describe a zero-frequency mode, which must have even parity  
by eq.(\ref{eq:A}). If $n$ is even, on the other hand, there is precisely one  
linearly independent zero-frequency mode associated with that integer, which  
may be written in the form
\begin{eqnarray}
      {\delta P(\vec{x})\choose\delta\rho(\vec{x})}
    =
const\frac{\rho_0(\vec{x})}{m}\left[\frac{V(\vec{x})}{k_BT}\right]^{n/2-1}{V(\vec{x})
\choose n/2}
\end{eqnarray}

The order of the secular equation and the number of frequencies with given
value of $n$ grows quadratically with $n$. For $n=0$ we have just one
coefficient $A_{000}$ and $P_1-\rho_1=0$, and the frequency $\omega^2=0$. We  
shall see in section \ref{subsec:d} that this mode must be excluded because of  
particle-number conservation. Thus, there is actually no physical mode with
$n=0$. For $n=1$ we have already three coefficients
$A_{100},A_{010},A_{001}$ but no B-coefficient yet. The equations for the
three A-coefficients are decoupled and we obtain the three frequencies
\begin{eqnarray}
      \omega_{100} = \omega_{x},
      \omega_{010} = \omega_{y},
      \omega_{001} = \omega_{z}
\label{eq:a}
\end{eqnarray}
corresponding to a rigid center of mass motion of the atom-cloud in the trap.  
These are the high-temperature limits of the Kohn-modes already encountered in
section \ref{subsec:2c}. Also the other temperature-independent modes
discussed there appear again as follows: For $n=2$ we have 6 A-coefficients  
and one
B-coefficient and all together 7 solutions for $\omega^2_{2}$. Because of
separate symmetry of the trapping potential under reflections of the x-, y-,  
and z-axis the three equations for
$A_{110},A_{101},A_{011}$  are decoupled from each other and the rest and
give immediately
\begin{eqnarray}
      \omega_{110} = \sqrt{\omega^2_x+\omega^2_y},
      \omega_{101} = \sqrt{\omega^2_x+\omega^2_z},
      \omega_{011} = \sqrt{\omega^2_y+\omega^2_z}.
\label{eq:b}
\end{eqnarray}
Similarly, for $n=3$ the coefficient $A_{111}$ is decoupled from the rest,
and the corresponding mode frequency is
\begin{eqnarray}
      \omega_{111} = \sqrt{\omega^2_x+\omega^2_y+\omega^2_z}.
\label{eq:c}
\end{eqnarray}
All these decoupled modes are isothermal modes, corresponding to
$\delta T=0$. In fact, we have already seen in subsection \ref{subsec:2c}
that these modes exist for all temperatures (for bosons for $T>T_c$).
The remaining four coefficients for $n=2, A_{200}, A_{020}, A_{002}$ and
$B_{000}$ are coupled. Of the four resulting coupled modes one is a
zero-frequency mode $\omega^{(0)}_2=0$, for which
\begin{eqnarray}
      \frac{1}{\omega^2_x}A^{(0)}_{200}
    = \frac{1}{\omega^2_y}A^{(0)}_{020}
    = \frac{1}{\omega^2_z}A^{(0)}_{002}
    = \frac{m}{2k_BT} B_{000}.
\label{eq:d}
\end{eqnarray}
This is the only internal-wave mode with $n=2$. The remaining three modes
with $n=2$ are modified sound modes, for whose squared frequencies
$\omega^2_2$
 the cubic equation.(\ref{eq:e}) is re-obtained as it must be, since the
corresponding modes were found in section \ref{subsec:2c} to exist for all  
temperatures with temperature-independent mode-frequencies (but with  
mode-functions for pressure and density which depend on temperature and the  
quantum-statistics).
The solutions of the cubic secular equation are simple if one trap frequency,  
say $\omega_z$, is much
smaller or much larger than the two others. Then one solution for $\omega^2$  
is small,
\begin{eqnarray}
      \omega\simeq \sqrt\frac{12}{5}\omega_z,
\label{eq:f}
\end{eqnarray}
or large
\begin{eqnarray}
      \omega\simeq \sqrt\frac{8}{3}\omega_z,
\label{eq:g}
\end{eqnarray}
respectively, while the other two are large and given by
\begin{eqnarray}
      \omega^2_{\pm}\simeq
      \frac{4}{3}
      \left(\omega^2_y+\omega^2_x\right)\pm\frac{4}{3}
      \sqrt{(\omega^2_y+\omega^2_x)^2-\frac{15}{4}\omega^2_y\omega^2_x}
\label{eq:h}
\end{eqnarray}
or small
\begin{eqnarray}
      \omega^2_{\pm}\simeq
      \frac{5}{4}
      \left(\omega^2_y+\omega^2_x\right)\pm\frac{5}{4}
      \sqrt{(\omega^2_y+\omega^2_x)^2-\frac{96}{25}\omega^2_y\omega^2_x},
\label{eq:i}
\end{eqnarray}
respectively. Other simple solutions are obtained in the isotropic case and
the axially symmetric case, cf. the
following sections. Quartic equations for $\omega^2$ similar to eq.(\ref{eq:e})  
are obtained from the solvability conditions for the three quadruples of
amplitudes of the
$n=3$ modes $(A_{102}, A_{120}, A_{300}, B_{100})$, $(A_{012}, A_{210},  
A_{030}, B_{010})$, and
$(A_{021}, A_{201}, A_{003}, B_{001})$, which decouple from each other  
because they
differ in parity. The solutions of these equations and of (\ref{eq:e}) for
arbitrary trap-frequencies are tedious expressions and will not be given here.  
They and the eigenvalues $\omega^2_n$ for $n\ge3$ can easily be determined
numerically for specific ratios of the trap-frequencies when needed, where
efficient use can be made of the already mentioned conservation of the x-, y-,  
and z-reflection parities.

\subsection{Isotropic case}\label{subsec:4b}

The solution for the isotropic case
$\omega_x=\omega_y=\omega_z=\omega_0$ in the high-temperature regime has
already been given by Bruun and Clark \cite{BC} based on the velocity equation  
(\ref{eq:2.8}). It is not clear, however, how the Hilbert space is defined in  
terms of the velocity variables. Indeed, the velocity-field turns out to be
polynomial like $\rho_1$ and $P_1$ and is therefore not a square-integrable
field. It is therefore worthwhile to check this case again, using our present  
description. Introducing dimensionless frequency and space-coordinate  via
\begin{eqnarray}
      \Omega^2 = \frac{\omega^2}{\omega^2_0},
      \vec{r} = \sqrt{\frac{m\omega^2_0}{k_BT}}\vec{x}
\label{eq:4.7}
\end{eqnarray}
and imposing a polynomial ansatz
\begin{eqnarray}
      P_1(\vec{r})
  &=& r^l Y_{lm}(\Theta,\varphi)Q_n(r^2)
\nonumber \\
      P_1(\vec{r})-\rho_1(\vec{r})
  &=& r^l Y_{lm}(\Theta,\varphi)T_{n-1}(r^2),
\label{eq:4.8}
\end{eqnarray}
with $Q_n$ and $T_n$ polynomials of order $n$ and $Y_{lm}(\Theta,\varphi)$
the spherical harmonics, we obtain

\begin{eqnarray}
      \left(\Omega^2-l-r\frac{d}{dr}\right) Q_n(r^2)
  &=& \left[\frac{5}{2}\Omega^2-r^2\right]T_{n-1}(r^2)
\label{eq:4.9}  \\
    - \frac{2}{3}\left(
      \frac{d^2}{dr^2}+\frac{2(l+1)}{r}\frac{d}{dr}\right)Q_n(r^2)
  &=& \left[\Omega^2-2-\frac{2l}{3}-\frac{2}{3}r\frac{d}{dr}
      \right]T_{n-1}(r^2).
\label{eq:4.10}
\end{eqnarray}
We should mention that the quantum-number $n$ defined by the polynomial order
of $Q_n, T_n$ differs from the quantum-number $n$ defined in section
\ref{subsec:4a}, which corresponds to $2n+l$ within the present definitions.
As was mentioned already, the characteristic equation follows from the linear  
set of equations for the largest powers. Let us consider separately

      (i) $\underline{n=0}$:
\begin{eqnarray}
          Q_0(r^2)=A^{(0)}_0, \qquad T_{-1}(r^2)=0
\label{eq:4.11}
\end{eqnarray}
Eq. (\ref{eq:4.10}) is consistent with $T_{-1}=0$ and gives no additional
information. From eq.(\ref{eq:4.9}) we conclude
\begin{eqnarray}
      (\Omega^2-l)A^{(0)}_0 = 0.
\label{eq:4.12}
\end{eqnarray}
Thus we have the spectrum
\begin{eqnarray}
      \omega = \sqrt{l}\omega_0
\label{eq:4.13}
\end{eqnarray}
for the modes
\begin{eqnarray}
      \delta P &=& \frac{k_BT}{m}\delta\rho
    =\frac{k_BT}{m}\rho_0(x)A^{(0)}_0 r^l Y_{lm}(\Theta,\varphi)
\nonumber \\
      \delta T
   &=& 0.
\label{eq:4.14}
\end{eqnarray}
These isothermal modes were first found by Griffin, Wu and Stringari \cite{GWS}

      (ii) $\underline{n\ge 1}$:
\begin{eqnarray}
           Q_n(r^2)
    &=&   \sum^{n}_{i=0}A^{(n)}_i r^{2i}
\nonumber\\
      T_{n-1}(r^2)
    &=& \sum^{n-1}_{i=0}B^{(n)}_i r^{2i}
\label{eq:4.15}
\end{eqnarray}
From (\ref{eq:4.9}) we obtain for
\begin{eqnarray}
        i
   =  n:&& \qquad (\Omega^2-l-2n)A^{(n)}_n +B^{(n)}_{n-1} = 0
\label{eq:4.16}   \\
        i
    \ne 0,n:&& \qquad (\Omega^2-l-2i)A^{(n)}_i + B^{(n)}_{i-1}
            -\frac{5}{2}\Omega^2 B^{(n)}_i = 0
\label{eq:4.17}  \\
        i
   =  0:&& \qquad (\Omega^2-l)A^{(n)}_0-\frac{5}{2}\Omega^2 B^{(n)}_0 = 0
\label{eq:4.18}
\end{eqnarray}
From (\ref{eq:4.10}) we get correspondingly
\begin{eqnarray}
      \frac{4}{3}i(2i+2l+1)A^{(n)}_i +
           \left[\Omega^2-\frac{2(l+1)}{3}-\frac{4i}{3}\right]
           B^{(n)}_{i-1} = 0
\label{eq:4.19}
\end{eqnarray}
Let us consider the highest power $i=n$ first, which yields two linear
homogeneous algebraic equations for $A^{(n)}_{n}, B^{(n)}_{n-1}$ of the form
\begin{eqnarray}
      \underline{\bf C}_{n-1}
                {A^{(n)}_{n} \choose B^{(n)}_{n-1}} = 0
\label{eq:4.20}
\end{eqnarray}
\begin{eqnarray}
      \underline{\bf C}_{n-1} =\left(
      \begin{array}{ccc}
      {\frac{4}{3}n\left[2n+2l+1\right]}
      &&
      {\Omega^2 - \frac{2(l+1)}{3} -\frac{4n}{3}}\\
      &&\\
      {\Omega^2-l-2n}
      &&
      {1}
      \end{array} \right)
\label{eq:4.21}
\end{eqnarray}
The solvability condition gives for $n\ge 1$
\begin{eqnarray}
      \Omega^4-\frac{5}{3}\Omega^2(l+2n+\frac{2}{5})
    + \frac{2}{3}l(l+1)=0
\label{eq:4.21a}
\end{eqnarray}
\begin{eqnarray}
      \Omega^2_{1,2} = \frac{\omega^2_{1,2}}{\omega^2_0}
    = \frac{1}{2}\left[
      \frac{5}{3}\left(l+2n+\frac{2}{5}\right)
    \pm \sqrt{\frac{25}{9}\left(l+2n+\frac{2}{5}\right)^2
    - \frac{8}{3}l(l+1)}\right]
\label{eq:4.22}
\end{eqnarray}
This result was first obtained by Bruun and Clark \cite{BC}. We
differ from their result only for the case $n=0$, where we find that only one  
branch of the
solution (\ref{eq:4.22}) exists, namely the branch which yields $\Omega^2=l$,  
as seen from (\ref{eq:4.11}), (\ref{eq:4.12}). In the case $n\ge 1, l=0$ one  
of the solutions (\ref{eq:4.22}) vanishes. It gives the zero-frequency
solutions discussed in previous sections. The other solution at
$\Omega^2=\omega^2_1/\omega^2_0=\frac{5}{3}(2n+\frac{2}{5})$ behaves normally  
and describes a sound-mode, as can e.g. be seen from the factor $5/3$ which
is characteristic of the inverse of the adiabatic compressibility in ideal
gases.

Let us turn to the mode-functions in the case $n\ge1$. Their coefficients can  
be obtained recursively starting with $B^{(n)}_0$, which is fixed only by
normalization, then solving (\ref{eq:4.18})
\begin{eqnarray}
      A^{(n)}_0 = \frac{5\Omega^2}{2\Omega^2-l}B^{(n)}_0,
\label{eq:4.23}
\end{eqnarray}
then rising to higher values of $i$ by first solving for
\begin{eqnarray}
      A^{(n)}_i = \frac{1}{4}
      \frac{3\Omega^2 -2(l+1)-4i}{i(2i+2l+1)} B^{(n)}_{i-1},
\label{eq:4.24}
\end{eqnarray}
and then using (\ref{eq:4.17}) with (\ref{eq:4.24}) and (\ref{eq:4.21a}) to
obtain also $B^{(n)}_i$ in terms of $B^{(n)}_{i-1}$ as
\begin{eqnarray}
      B^{(n)}_i = - \frac{n-i}{i(2i+2l+1)}B^{(n)}_{i-1}.
\label{eq:4.25}
\end{eqnarray}
Solving the recursion-relation (\ref{eq:4.25}) we obtain
\begin{eqnarray}
      B^{(n)}_i = \frac{(1-n)(1-n+1)\cdots(1-n+i-1)}
      {(\frac{2l+3}{2})\cdot(\frac{2l+3}{2}+1)\cdots(\frac{2l+3}{2}+i-1)}
      \frac{1}{i!2^i} B^{(n)}_{0}
\label{eq:4.26}
\end{eqnarray}
which identifies the polynomial $T_{n-1}(r^2)$ as the confluent
hypergeometric function
\begin{eqnarray}
      T_{n-1}(r^2) = \sum^{n-1}_{i=0} B^{(n)}_i r^{2i}
    = B^{(n)}_{0} {}_1F_1(1-n,\frac{2l+3}{2}; \frac{r^2}{2})
\label{eq:4.27}
\end{eqnarray}
which is proportional to the generalized Laguerre-polynomial
$L^{(l+\frac{1}{2})}_{n-1}(\frac{r^2}{2})$. Substituting the result into
eq.(\ref{eq:4.24}) we obtain also the polynomial $Q_n(r^2)$ as the combination  
of hypergeometric functions
\begin{eqnarray}
      Q_n(r^2) = \sum^{n}_{i=0} A^{(n)}_i r^{2_i} =B^{(n)}_{0} && \left\{
      \frac{3}{4n}\left[\Omega^2_{1,2}+\frac{2l}{3}\right]
      {}_1F_1\left(-n, \frac{2l+3}{2};\frac{r^2}{2}\right) \right.\nonumber\\
       &&\left. -\frac
      {2l+1}{2n} {}_1F_1 \left(-n, \frac{2l+1}{2};\frac{r^2}{2}\right)
      \right\}
\label{eq:4.28}
\end{eqnarray}
which is proportional to
$(3\Omega^2_{1,2}+2l)L^{(l+\frac{1}{2})}_n (\frac{r^2}{2})
-2(1+2l+2n)L^{(l-\frac{1}{2})}_n(\frac{r^2}{2})$. The physical modes are then
\begin{eqnarray}
      \delta P (\vec{x},t)
   &=& \frac{k_BT}{m}\rho_0(\vec{x})r^lY_{lm}(\Theta,\varphi)Q_n(r^2)
\label{eq:4.29}   \\
      \frac{\delta T (\vec{x},t)}{T}
   &=& r^l Y_{lm}(\Theta,\varphi)T_{n-1}(r^2)
\label{eq:4.30}
\end{eqnarray}
It is quite remarkable that the spatial perturbation of the temperature in
the two physically very different branches of the spectrum, the sound-modes
and the internal modes, is exactly the same, because $T_{n-1}(r^2)$ is
independent of $\omega^2_{\pm}$ and therefore the same for both branches.
(We note, however, that this is strictly true only in the Boltzmann limit). The  
spatial distribution of the pressure and the density, on the other hand, is  
very different for
both kinds of modes as one would expect.

For $l=0$ and arbitrary $n$ we obtain the mode-functions of the
zero-frequency modes. They form the bottom of a ladder of rotational modes for  
each value of $n$. The mode-functions are given by combinations of
Hermite-polynomials of the radial variable $(m\omega^2_0\vec{x}^2/k_BT)^{1/2}$  
and form a complete set in the Hilbert-space of radial functions defined by
the scalar product (\ref{eq:4.6a}). This simply means that an
\underline{arbitrary} radial, i.e. angle-independent, mode-function within our  
Hilbert-space is a zero-frequency mode.

\subsection{Anisotropic harmonic potential and  conserved operator}
            \label{subsec:4c}

Let us now turn to the case of an anisotropic harmonic trapping potential.
Then the generator of rotations $\underline{\makebox{\bf
1}}\vec{L}=-i\underline{\makebox{\bf 1}}(\vec{x}\times\vec{\nabla})$ no longer  
commutes with the wave-operator $\underline{\makebox{\bf H}}$ defined by the  
matrix-differential operator on the right-hand side of (\ref{eq:4.5}). On the  
other hand, for harmonic trapping potentials the polynomial solutions
discussed in section \ref{subsec:4a} continue to exist also in this case.
Therefore, one must strongly suspect that a complete set of operators
commuting with $\underline{\makebox{\bf H}}$ exists also in the fully
anisotropic case. For Bose-Einstein condensed bosons and for fermions at
temperature $T=0$ a similar situation prevails (but not at temperatures
between the high- and the low-temperature limits, cf. the discussion below),  
and two operators commuting with the corresponding wave-operator were
constructed in our previous papers \cite{CG}, \cite{CG2}. One way to find
these operators is to introduce elliptic coordinates and to separate the
wave-equation in these coordinates. The separation-constants introduced by
this procedure appear naturally as eigenvalues of certain differential
operators, any combination of which can be identified with the searched for
commuting operators. In \cite{CG}, \cite{CG2} this procedure was fully carried  
through in the low-temperature limit.

For our present high-temperature regime we shall not present the most general  
anisotropic case. Rather we restrict ourselves to axially symmetric traps
with $\omega_x=\omega_y=\omega_{\perp}$, where one of the conserved
operators, namely $\underline{\makebox{\bf
1}}L_z$ with
\begin{eqnarray}
      L_z = -i\left(x \frac{\partial}{\partial y}- y \frac{\partial}{\partial  
x}\right),
 \label{4.31}
\end{eqnarray}
still follows from symmetry and we only need to find a second one to
demonstrate integrability.

One can find a conserving operator via the
separation of the wave equation for $P_1$ in elliptical cylindric coordinates.
This procedure leads to an operator $\underline{\makebox{\bf B}}'$, which has the
important property
\begin{eqnarray}
      \underline{\makebox{\bf B}}'{P_1\choose P_1-\rho_1} = B
{P_1\choose P_1-\rho_1},
\label{eq:4.35}
\end{eqnarray}
where the eigenvalue $B$ is the separation constant.
$\underline{\makebox{\bf B}}'$ has the form
\begin{eqnarray}
      \underline{\makebox{\bf B}}' = \left(
      \begin{array}{ccc}
      {\hat{R}-\frac{4}{5}}
      && {2} \\
      {0}
      && {\hat{R} +\frac{6}{5}}
      \end{array}     \right),
\label{eq:4.36}
\end{eqnarray}
where
(with
$\omega^2_x=\omega^2_y=\omega_\perp^2$)
\begin{eqnarray}
      \hat{R}
    = \vec{x}\cdot\vec{\nabla}-\frac{k_BT}{m}\left(
      \frac{1}{\omega^2_x}\frac{\partial^2}{\partial x^2}
      + \frac{1}{\omega^2_y}\frac{\partial^2}{\partial y^2}
      + \frac{1}{\omega^2_z}\frac{\partial^2}{\partial z^2}\right).
\label{eq:4.38}
\end{eqnarray}
Returning back to our original variables $\delta P$ and $\delta \rho$
one can prove via explicit calculation
that the operator $\underline{\makebox{\bf B}}$, which
corresponds to $\underline{\makebox{\bf B}}'$ in that representation,
has vanishing commutator with $\underline{\makebox{\bf H}}$:
\begin{eqnarray}
      \left[\underline{\makebox{\bf H}},\underline{\makebox{\bf B}}\right]
    = 0
\label{eq:4.39}
\end{eqnarray}

Let us determine the spectrum of $\underline{\makebox{\bf B}}$ in the
Hilbert-space of polynomials discussed in section \ref{subsec:4a}, where $P_1$  
is a polynomial of total order $n$ and $P_1-\rho_1$ a polynomial of total
order $n-2$. We obtain directly from (\ref{eq:4.35}), (\ref{eq:4.38}) the
eigenvalue
\begin{eqnarray}
       \makebox{B} = n - \frac{4}{5}
\label{eq:4.40}
\end{eqnarray}
i.e. $\underline{\makebox{\bf B}}$ is the conserved operator which introduces  
the principal quantum number $n$ of section \ref{subsec:4a} which is the
total number of nodal surfaces of the pressure and density oscillations in any  
given mode.

\subsection{Restrictions by particle-number conservation}
            \label{subsec:d}
In the Hilbert-space defined by the scalar product (\ref{eq:2.30}) there may  
still be unphysical states, which do not satisfy the conservation of the total  
number of particles
\begin{eqnarray}
      N = \frac{1}{m}\int d^3x\left(
          \rho_0(\vec{x})+\delta\rho(\vec{x},t)\right)
\label{eq:4.41}
\end{eqnarray}
which applies to a closed system. In terms of the variable
$\rho_1=\delta\rho/\rho_0$ eq.(\ref{eq:4.41}), with
$N=\frac{1}{m}\int d^3x\rho_0(\vec{x})$ and
$\rho_0(\vec{x})=\rho_0(0)e^{-V(\vec{x})/k_BT}$, implies the condition
\begin{eqnarray}
      \int d^3x e^{-V(\vec{x})/k_BT} \rho_1 = 0
\label{eq:4.42}
\end{eqnarray}
In order to see which modes satisfy (\ref{eq:4.42}) we use the existence of
the conserved $\underline{\makebox{\bf B}}$ with (\ref{eq:4.36}), from which,  
by elimination of $P_1$, the eigenvalue equation for $\rho_1$
\begin{eqnarray}
      B\rho_1 = \left(
          -\frac{4}{5}+\hat{R}\right)\rho_1
\label{eq:4.43}
\end{eqnarray}
can be derived very easily.
We can use this to insert $\hat{R}$ in front of $\rho_1$ in
eq.(\ref{eq:4.42}) to obtain
\begin{eqnarray}
       \int d^3x e^{-V(\vec{x})/k_BT}\hat{R}\rho_1
     = \left( B+\frac{4}{5}\right)\int d^3x e^{-V(\vec{x})/k_BT}\rho_1
\label{eq:4.44}
\end{eqnarray}
Now we use the explicit form of $\hat{R}$ and the quadratic form of the
potential on the left hand side of (\ref{eq:4.44}) and apply partial
integration to let the derivatives act on the exponential factor. Boundary
terms are not incurred by this operation because $\rho_1$ is polynomial and
the exponential factor vanishes rapidly at infinity. We find that the left
hand side of (\ref{eq:4.44}) vanishes, and therefore
\begin{eqnarray}
      \left( B+\frac{4}{5}\right)\int d^3x e^{-V(\vec{x})/k_BT}
      \rho_1= 0
\label{eq:4.45}
\end{eqnarray}
Hence, all modes for which the eigenvalue $B$ satisfies
\begin{eqnarray}
      B \ne -\frac{4}{5}
\label{eq:4.46}
\end{eqnarray}
satisfy the restriction (\ref{eq:4.42}) imposed by particle-number
conservation. Looking at the spectrum (\ref{eq:4.40}) of
$\underline{\makebox{\bf B}}$ we see that (\ref{eq:4.46}) is satisfied for all  
modes with the exception of the mode with $n=0$, which is a special
zero-frequency mode. For this mode $\rho_1(\vec{x})=const$ and (\ref{eq:4.42})  
is obviously violated. This mode must therefore be excluded from the physical  
spectrum. All the other modes, and in particular all the other zero-frequency  
modes, satisfy particle-number conservation.

\subsection{Solutions for axially symmetric traps}
            \label{subsec:e}
We introduce scaled cylinder coordinates $r,z,\varphi$ via
\begin{eqnarray}
      x_1=\sqrt{\frac{k_BT}{M\omega^2_{\perp}}}r\cos\varphi,\qquad
      x_2=\sqrt{\frac{k_BT}{M\omega^2_{\perp}}}r\sin\varphi,\qquad
      x_3=\sqrt{\frac{k_BT}{M\omega^2_z}}z
\label{eq:4.47}
\end{eqnarray}
and the ansatz
\begin{eqnarray}
      P_1= r^{|m|}e^{im\varphi}z^{\alpha} Q(\rho^2,z^2)
\nonumber \\
      P_1-\rho_1=r^{|m|}e^{im\varphi}z^{\alpha} T(\rho^2,z^2)
\label{eq:4.48}
\end{eqnarray}
where $\alpha =0,1$ determines the parity under inversion of the z-axis and
$m=-\infty,\cdots, 0,1,2,\cdots\infty$ is the quantum number of angular
momentum around the z-axis. The equations to be solved are then, with
$\Omega^2=\omega^2/\omega^2_{\perp}, \lambda=\omega^2_z/\omega^2_{\perp}$
\begin{eqnarray}
      \Omega^2 T
  &=& -\frac{2}{3}\left[
      \frac{\partial^2}{\partial\rho^2} + \frac{\partial^2}{\partial z^2}
    + \frac{2|m|+1}{\rho}\frac{\partial}{\partial\rho}
    + \frac{2\alpha}{z}\frac{\partial}{\partial z}\right] Q
\nonumber \\
 && + \frac{2}{3}\left[
      |m|+2+\lambda(\alpha+1)+\rho\frac{\partial}{\partial\rho}
    + \lambda z \frac{\partial}{\partial z}\right] T
\label{eq:4.49}  \\
      \Omega^2 Q
  &=& \left[
      \frac{5}{2}\Omega^2-\rho^2-\lambda^2z^2\right] T
    + \left[ |m|+\lambda\alpha+\rho\frac{\partial}{\partial\rho}
    + \lambda z\frac{\partial}{\partial z}\right] Q.
\nonumber
\end{eqnarray}
For $Q$ and $T$ we make a polynomial ansatz in $\rho^2,z^2$ of order $n$ and  
$n-1$, respectively. The eigenvalue of
$\underline{\makebox{\bf B}}'$ for these
solutions is $B=2n+|m|+\alpha-\frac{4}{5}$. The terms of $Q$ and $T$ of
highest order read
\begin{eqnarray}
      Q&=& \sum^{n}_{l=0}A_l \rho^{2l}z^{2(n-l)}
                                  \qquad\makebox{+ lower order}
\nonumber \\
      T&=&\sum^{n-1}_{l=0} B_l\rho^{2l}z^{2(n-1-l)}
                                  \qquad\makebox{+ lower order}
\label{eq:4.50}
\end{eqnarray}
Inserting the ansatz and comparing coefficients of the highest order terms we  
obtain the two sets of equations
\begin{eqnarray}
      \Omega^2 & A_0& =
      \left[|m|+\lambda(2n+\alpha)\right]A_0-\lambda^2B_0
\nonumber \\
      \Omega^2 & A_1 &=
            - B_0+\left[|m|+2+\lambda(2n-2+\alpha)\right]A_1-\lambda^2B_1
\label{eq:4.51} \\
            &\vdots&
\nonumber \\
      \Omega^2 & A_l & =
      -B_{l-1}+\left[|m|+2l+\lambda(2n-2l+\alpha)\right]A_l-\lambda^2B_1
\nonumber\\
            &\vdots&
\nonumber  \\
      \Omega^2  &A_n & =
      -B_{n-1}+\left[|m|+2n+\lambda\alpha\right]A_n
\nonumber
\end{eqnarray}
and
\begin{eqnarray}
      \Omega^2 & B_l = &
      -\frac{4}{3}(n-l)(2n-2l-1+2\alpha)A_l
\nonumber\\
&&    +\frac{2}{3}\left[
       |m|+2l+2+\lambda(2n-2l+\alpha-1)\right]B_l
\label{eq:4.52}  \\
&&     -\frac{8}{3}(l+1)(|m|+l+1)A_{l+1}
\nonumber
\end{eqnarray}
Let us examine simple special cases.

For $n=0$ we have $Q=A_0, T=0$ and obtain
\begin{eqnarray}
      \Omega^2 = \frac{\omega^2}{\omega^2_{\perp}}
    =  |m|+\lambda\alpha
      \qquad (|m|=0,1,\cdots; \alpha=0,1)
\label{eq:4.53}
\end{eqnarray}
This result is a special case of the result (\ref{eq:a}) (for $m=0,\alpha=1$  
and $m=\pm 1,\alpha=0$), and (\ref{eq:b}) (for $m=\pm 1,\alpha=1$ and one mode  
with $|m|=2,\alpha=0$), and (\ref{eq:c}) (for one mode with
$|m|=2,\alpha=1$), but the case $|m|\ge3$ has no simple counterpart in the
fully anisotropic case.

For $n=1$ we get already the three coupled equations
\begin{eqnarray}
      \lambda^2B_0
   &+& [\Omega^2-|m|-(2+\alpha)\lambda]A_0   =0
\nonumber \\
      B_0
   &+& (\Omega^2-|m|-2-\lambda\alpha)A_1 =0
\label{eq:4.54} \\
      \left[\Omega^2-\frac{2}{3}(|m|+2+\lambda+\alpha\lambda)\right]
       B_0
   &+&\frac{4}{3}(1+2\alpha)A_0+\frac{8}{3}(|m|+1)A_1  =0
\nonumber
\end{eqnarray}
Specializing further to $m=0$ and $\alpha=0$ we find the exact solutions
\begin{eqnarray}
      \omega^2_1 =0  \qquad\makebox{with}\qquad
      A_0=\lambda A_1=\frac{\lambda}{2}B_0
\label{eq:4.55}
\end{eqnarray}
and
\begin{eqnarray}
      \omega^2_{2,3} =
      \frac{\omega^2_{\perp}}{3}\left[
      4\lambda+5\pm\sqrt{16(\lambda-1)^2+9}\right].
\label{eq:4.56}
\end{eqnarray}
These mode-frequencies were first obtained in \cite{GWS}.

The zero-frequency mode is the axially symmetric counterpart of (\ref{eq:d})  
while (\ref{eq:4.56}) corresponds to two of the three mode-frequencies which  
solve (\ref{eq:e}). The third solution of (\ref{eq:e}) in the axially
symmetric case is $\omega^2=2\omega^2_{\perp}$
and is in the completely  
anisotropic case the counterpart of the second of the two modes described by  
(\ref{eq:4.53}) with $|m|=2, \alpha=0$.

In order to account for all modes in our comparison between the axially
symmetric and the fully anisotropic case, we may finally note that the second  
of the two modes with $|m|=2, \alpha=1$ described by (\ref{eq:4.53}) has as a  
completely anisotropic counterpart a particular solution of the quartic
secular-equation for $\omega^2$ for the amplitudes $(A_{021}, A_{201},
A_{003},
B_{001})$ appearing in the scheme of section \ref{subsec:4a} for $n=3$.

\section{Numerical determination of the temperature-dependent
         mode-spectrum}\label{sec:5}
At intermediate temperatures the coupled wave equations (\ref{eq:2.13}),
(\ref{eq:2.14}) are generally
not separable and the spectrum can only be found
numerically. Best suited for numerical work are the wave equations in the
manifestly hermitian form (\ref{eq:2.44}) with the scalar product in the
simple form (\ref{eq:2.39}).

\begin{figure}
\centerline{\epsfig{figure=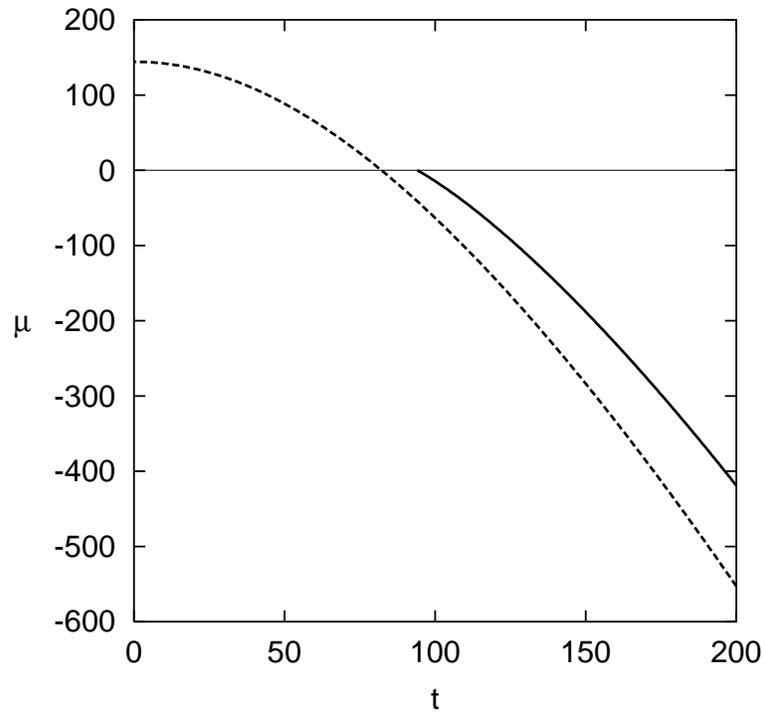,width=10.cm,angle=0}}
\caption{The dimensionless chemical potential $\mu/\hbar\bar{\omega}$
in the range $\mu<0$ as a function of the dimensionless temperature  
$t=k_BT/\hbar\bar{\omega}$  with $\bar\omega=(\omega_z\omega_\perp^2)^{1/3}$,
for bosons (full curve) and for fermions
(dashed curve); number of atoms $N=10^6$.}
\label{fig:2}
\end{figure}

The numerical analysis will be performed for the axially symmetric case,
choosing for the anisotropy parameter $\lambda=\omega^2_z/\omega^2_{\perp}=8$,  
partially for historical reasons as this was the geometry of the first TOP
trap at JILA \cite{TOP}. The number of particles is chosen as $N=10^6$. First  
the chemical potential is determined for the given particle-number $N$ as a
function of temperature. This is done in the standard way, by integrating
eq.(\ref{eq:1.4}) to obtain $N(\mu,T)$ and solving for $\mu(N,T)$. The results  
for our chosen set of parameters are displayed in fig.\ref{fig:2} both for
bosons and for fermions, in the domain where $\mu\le 0$, to which we shall
restrict our attention, in the following. We may remark here that it follows  
from the form of the potential and (\ref{eq:1.4}) that $\mu$ is a scaling
function $S$ of $N,\bar{\omega}=(\omega_x \omega_y \omega_z)^{1/3}$, and $T$  
of the form $\mu/\hbar\bar{\omega}=S(k_BT/\hbar\bar{\omega}N^{1/3})$.

In order to determine the spectrum, the wave-operator (\ref{eq:2.44}) is
represented in the basis of the harmonic oscillator eigenfunctions
with widths $\sqrt{k_BT/m\omega_z^2}$ and $\sqrt{k_BT/m\omega_\perp^2}$ in  
axial and radial direction, respectively. The basis is cut-off at a
finite size of order 100 both for $u$ and $v$ and the resulting  
finite-dimensional Hermitian matrix is
diagonalized.
The size of the finite basis is varied in order to control that
the eigenvalues obtained are converged numerically.
The truncation of the basis to a finite size introduces some spurious  
eigenmodes and eigenfrequencies, which can be distinguished however, and  
subsequently eliminated, by the fact that they don't converge
but disappear and reappear somewhere else as the size of the basis is varied.

\begin{figure}
\centerline{\epsfig{figure=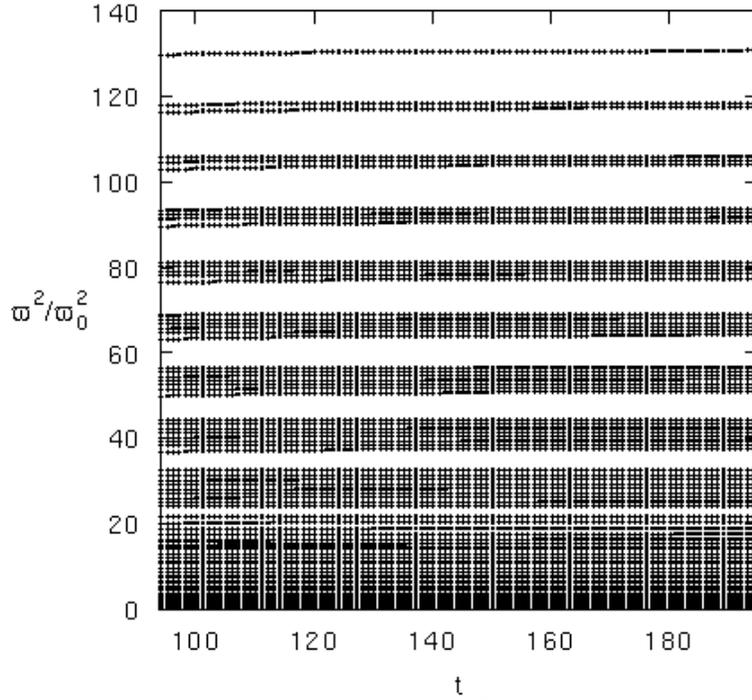,width=10.cm,angle=0}}
\caption{Dimensionless squared hydrodynamic mode-frequencies  
$(\omega/\bar{\omega})^2$ of a bosonic gas above the
BEC-transition
as function of the dimensionless temperature t, for the m=0 modes of even  
parity with up to 10 nodal surfaces; anisotropy parameter
$\lambda=(\omega_z/\omega_\perp)^2=8$; number of atoms $N=10^6$; total size  
of basis
132.}
\label{fig:3}
\end{figure}

\begin{figure}
\centerline{\epsfig{figure=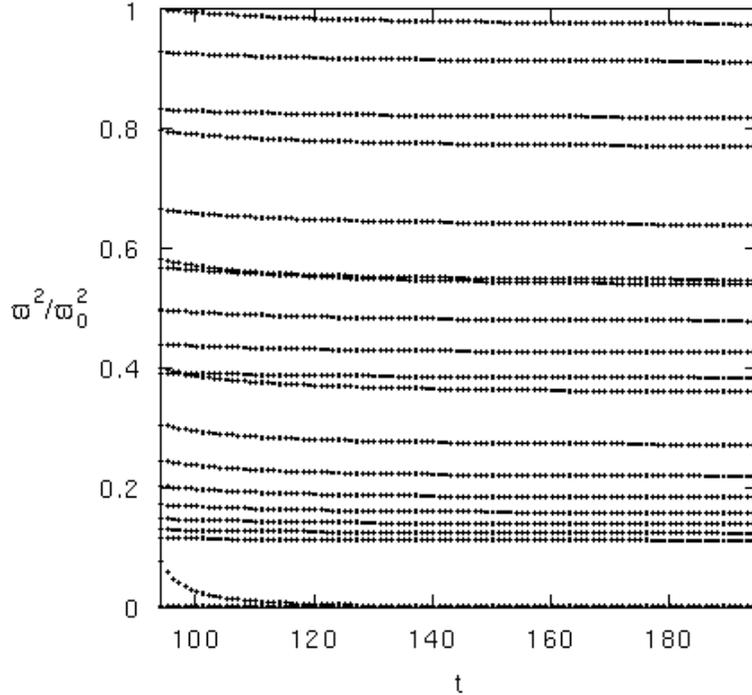,width=10.cm,angle=0}}
\caption{Squared internal mode frequencies of fig.\ref{fig:3} below the
geometric mean trap frequency as function of temperature, for the same
parameters as in fig.\ref{fig:3}.}
\label{fig:4}
\end{figure}

\begin{figure}
\centerline{\epsfig{figure=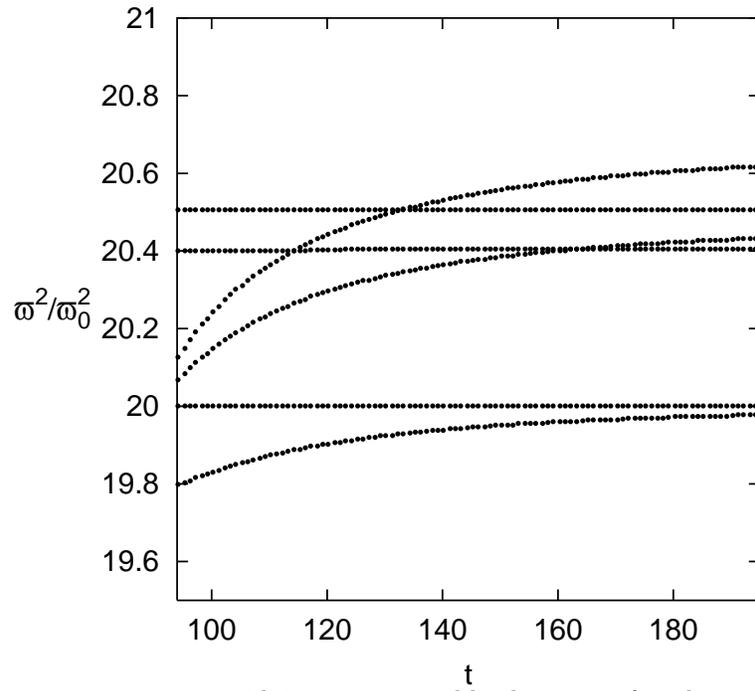,width=10.cm,angle=0}}
\caption{Squared mode-frequencies increasing with temperature and
level crossings for a bose-gas with the same parameters as in fig.\ref{fig:3},
for an isotropic trap.}
\label{fig:5}
\end{figure}

\begin{figure}
\centerline{\epsfig{figure=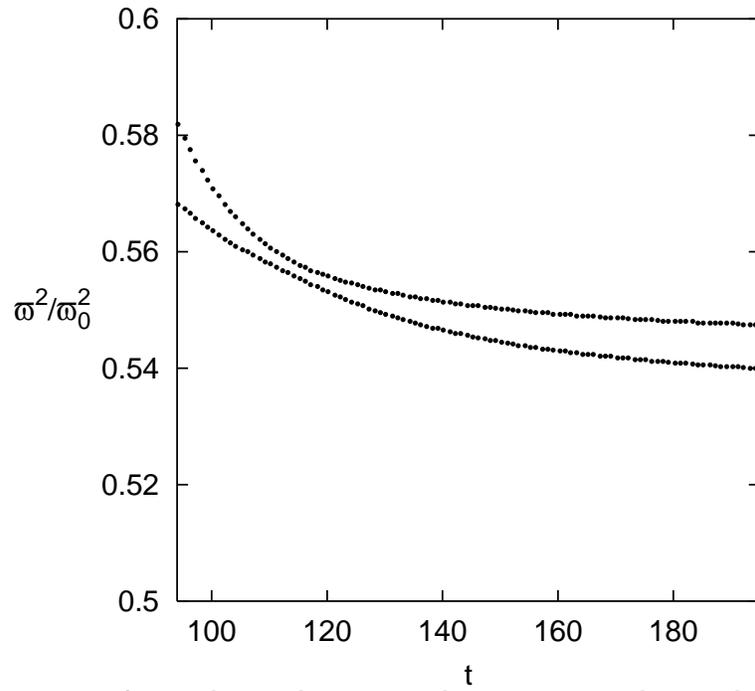,width=10.cm,angle=0}}
\caption{Squared mode-frequencies of internal waves decreasing with
temperature, and an avoided
level crossing for a bose-gas in an axially symmetric trap with the same
parameters as in fig.\ref{fig:3}.}
\label{fig:6}
\end{figure}

\begin{figure}
\centerline{\epsfig{figure=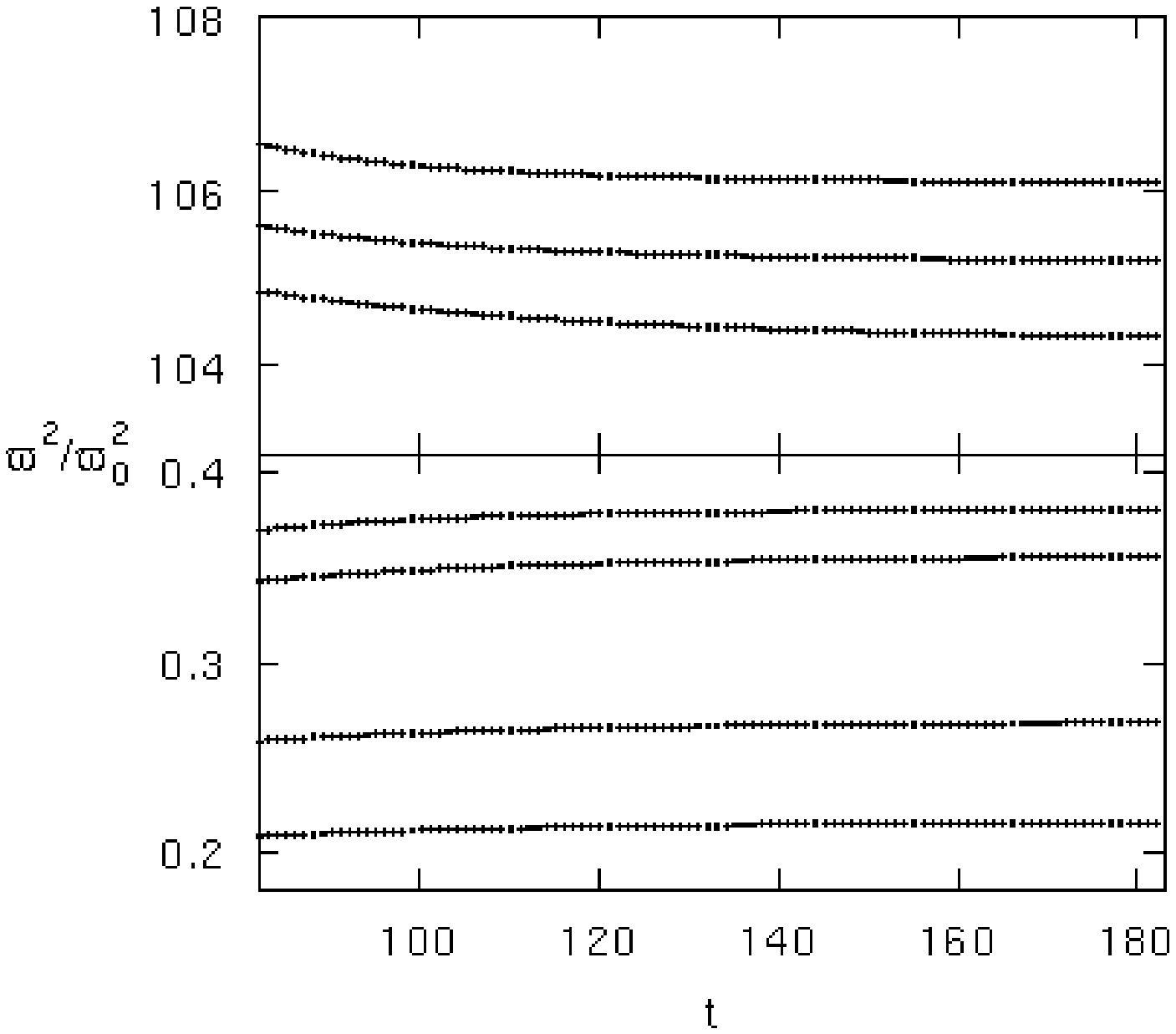,width=10.cm,angle=0}}
\caption{Squared mode-frequencies for a fermionic gas above
the Fermi-Temperature
as a function of $t=k_BT/\hbar\bar{\omega}$ for sound modes (upper part)
and for internal wave modes (lower part).
}
\label{fig:7}
\end{figure}

Some of the results are displayed in figs.\ref{fig:3}-\ref{fig:7} which we
now discuss. Fig.\ref{fig:3} gives an overview, in the domain $94<
k_BT/\hbar\bar{\omega}<195$, of the spectrum of eigenvalues $\omega^2$, for a  
gas of bosons (but very similar results not shown here are also found for  
fermions). For clarity only modes with azimuthal quantum number $m=0$ and
even parity are shown. The basis used consisted of  oscillator eigenfunctions  
of order $2n_z$
in z-direction and order $2n_\rho$ in radial direction, with integer
$n_\rho+n_z\le 10$. All  frequencies obtained by the diagonalization of the
matrix of $\underline{\makebox{\bf H}}$ in this basis (except
for the spurious ones
introduced by the truncation of the basis) are shown in the figure.
As can be seen from fig.\ref{fig:2} the chosen temperature domain extends
from the BEC-temperature to the high-temperature region. In fact, we have
checked that our numerical code applied to the case of an isotropic trap gives  
a similar result, which for $k_BT/\hbar\bar{\omega}=195$ coincides with the
analytically known eigenvalues (\ref{eq:4.22}) in the high-temperature limit  
with high precision. This fact offers us the possibility to assign the quantum  
numbers of the high-temperature domain to the whole corresponding
temperature-dependent branch of frequencies.

An obvious feature of fig.\ref{fig:3} is the band-like structure of the
spectrum, which is caused by the anisotropy of the trap: Because of the higher  
stiffness of the trap in axial direction for the assumed value of $\lambda=8$,  
nodes of the mode-function in axial direction are more costly in energy
$\hbar\omega$, for sound-like modes, than nodes in radial direction.
Therefore we can assign to the `bands' the quantum number $n_z$ of nodes of
sound-like modes in axial direction with $n_z=10$ in the highest `band' (which  
can consist of a single mode only in the subspace we consider in
fig.\ref{fig:3}) and $n_z=0$ in the lowest. It should be noted that the two
lowest bands are not split and form a single broad band. Within a given band  
sound-like modes differ only by the radial quantum number $n_\rho$ counting
the number of nodes in radial direction. By our choice of a finite-dimensional  
basis we are restricted to modes with quantum numbers $n_\rho+n_z\le 10$. The  
presence of internal waves complicates the assignment of quantum numbers,
because each pair of quantum numbers $n_z,n_\rho$ appears twice, once at
higher frequency for a sound-like mode and once at lower frequency for an
internal wave mode. Thus internal modes with high values of $n_z$ give also
frequencies in the low lying bands of fig.\ref{fig:3}. The number of  
eigenfrequencies in the 'bands' depends, of course, on the size of the
basis.

By these considerations we arrive at the following assignment of quantum
numbers to the frequencies shown in fig.\ref{fig:3}. The mode with the largest  
frequency forms the `band' on the top and must be a sound-mode with $n_z=10,  
n_\rho=0$. The next lower band must contain two sound-like modes with $n_z=9,  
n_\rho=1$ and $n_z=9, n_\rho=0$.
We see from the figure that this exhausts already the number of modes (=2) in  
the second 'band' from the top, which can therefore not contain an internal
wave mode. Similarly, in the $3^{rd}, 4^{th}, 5^{th}$, etc. 'band' from the
top there are $3, 4, 5$, etc. sound-like modes with $n_z=8, 7, 6$, etc. and
$3, 4, 5$, etc. different values of $n_\rho$, respectively. In the figure we  
can follow this counting of different frequencies down to the third 'band'
from the bottom, thereby accounting for all eigenvalues in these 'bands'. It  
follows that none of these 'bands' contains any internal wave modes.

The lowest two bands, on the other hand, in particular the lowest one,
contain many more different frequencies
than the 10 and 11 different sound-frequencies with $n_z=1, n_z=0$,
respectively. These must therefore be considered as the internal waves.
Among the internal waves are also the zero-frequency modes. For each value of
$n=n_z+n_\rho\ge1$ in our subspace there is precisely 1 zero-frequency mode,
so the eigenvalue $\omega^2=0$ is 10-fold degenerate in our subspace.
By their special nature the zero-frequency modes have quantum-numbers
$n_z=n_\rho$. Internal modes of high $n=2(n_z+n_\rho)$, for which $n_\rho$
differs only slightly from $n_z$ have frequencies close to zero

The discussion of the spectrum we have given depends on our arbitrary
restriction of the number of nodal surfaces to a finite and not very large
number. If modes with an arbitrary number of nodal surfaces are permitted, then  
the mode-spectrum becomes dense, due to the existence of small internal wave  
frequencies from modes with arbitrarily high quantum-numbers. This can already  
be seen from the analytically determined spectrum (\ref{eq:4.22}) for the
isotropic case.

Within the subspace of Hilbert space we consider here there are many internal  
mode frequencies also below the geometric mean trap frequency, as can be seen  
with more clarity in fig.\ref{fig:4}. Only a small number of
sound-modes can occur in this
regime. The mode frequencies displayed in figs.\ref{fig:3}, \ref{fig:4} have a  
surprisingly weak temperature-dependence throughout the range
considered. For large $T$ this can also be seen from the analytical results.  
For the sound-modes the velocity of sound increases with temperature roughly  
$\sim\sqrt{k_BT}$, but the wavelength of any given mode also increases with
temperature $\sim\sqrt{k_BT}$ due to the expansion of the size of the thermal  
cloud, so that both temperature-dependences effectively cancel. However, for  
internal modes the compensation between the speed of sound and the wave-length  
does not work in the same way, as can e.g. be seen from eq.(\ref{eq:3.7}). In  
fig.\ref{fig:5} we consider a magnification of a part of the frequency
spectrum (in this instance for the isotropic case) where the increase of
some frequencies with temperature can be seen, which freely cross other
levels  (which must have therefore different quantum numbers $n,l$), which
are nearly temperature-independent and belong to sound-like modes. However, in  
the axially symmetric case avoided crossings can also be seen, as shown in
fig.\ref{fig:6}
for two internal wave modes at small frequency below the geometric mean
trap-frequency. This indicates that the conserved operator
$\underline{\makebox{\bf B}}$ of section \ref{subsec:4c}
ceases to exist at intermediate temperatures. It would be futile, therefore,
 to look for analytical solutions of the spectrum in the intermediate
temperature range, as the system appears to be non-integrable.

Generally, the differences between the results for fermions and bosons
in the region above the degeneracy temperature are qualitatively
not very big. One difference due to quantum statistics can be seen in
fig.\ref{fig:7}, where the temperature dependence of internal wave modes
and sound modes is displayed for a Fermi gas. The frequencies of
the internal wave modes
curve downwards, those of the sound modes curve upwards. In the Bose gas
case the opposite tendency is found as shown in figures \ref{fig:4} and
\ref{fig:5}.

\section{Conclusions}\label{sec:6}

In the present paper we have given a systematic analysis of the hydrodynamic  
modes of quantum gases in a harmonic trap with general anisotropy
in the collision-dominated non-dissipative limit. Provided the hydrodynamic  
limit is applicable the analysis applies for Fermi-gases
at all temperatures and for Bose-gases at temperatures above
Bose-Einstein condensation. Our results extend previous works by allowing for  
traps with arbitrary anisotropy, and by treating bosons and fermions
side by side on an equal footing. In addition to analytical solutions in
certain special cases and limits we also present numerical solutions
in the whole temperature-domain covered by the theory. Our analysis is  
based on the reduction of the five conservation-laws for the densities of  
mass, momentum and energy to two coupled wave-equations for the mass-density  
and the pressure. We have constructed a scalar product with a positive  
$L_2$-norm on the space of solutions, in which the two-component wave-operator
is hermitian and, because of the stability of the hydrodynamic modes
which we demonstrate, it is non-negative. However, a class of solutions with
vanishing frequencies was found which is a consequence of the existence of
mechanical equilibrium states in addition to the unique thermodynamic  
equilibrium which maximizes the entropy. A further class of exact
solutions consisting of isothermal modes was identified among which
are the center of mass modes required by the Kohn-theorem.
These results generalize earlier results by Griffin et al. \cite{GWS}
 and Bruun and Clark \cite{BC} to traps without axial symmetry.

We studied the two coupled wave equations for pressure and density in the
short-wavelength limit. Two different branches of solutions could be  
identified in this way, the high-frequency branch being associated with  
pressure-driven sound waves,
the low-frequency branch with potential-driven internal waves. The explicit  
dispersion relation of the lower branch found in the short wavelength limit
and its characteristic properties like the existence of a maximal frequency,
the anisotropy of the dispersion-relation and the orthogonality of the local  
group velocity and the local wave-vector makes the identification
with internal waves manifest and unambiguous. We also examined the  
high-temperature limit of the two coupled wave equations and demonstrated  
the existence of polynomial solutions by exhibiting a  
conserved operator whose eigenvalues fix the respective polynomial order,
or, equivalently, the number of nodal surfaces of the solutions.
We constructed the solutions for density and pressure also explicitly.

The wave equations were finally also solved numerically. Surprisingly, the  
mode-spectrum found turned out to be quasi-continuous, which can be understood  
by the overlap of the spectrum of low-frequency internal waves of short  
wavelengths with the spectrum of sound waves with large wavelengths.
Within a finite-dimensional subspace of the Hilbert space, defined by
restricting from above the number of nodal surfaces, we found a band-like  
structure of the eigenfrequencies in a strongly anisotropic trap, where the  
bands are labeled by the number of nodal surfaces orthogonal to the more  
strongly confined direction while the modes within a given band differ
by the number of nodes in the other directions. The number of internal wave  
modes  in such a finite-dimensional subspace is also limited, and these modes  
are then found primarily in the lowest lying band. We note that the  
restriction to a finite subspace of Hilbert space is also physically  
motivated, since
typical excitation-mechanisms like the modulation of the trapping potential,  
will also excite only modes in a certain subspace with appreciable amplitude.

The analysis we have presented is subject to some obvious limitations
which we now discuss briefly. It is clear that only systems in the  
collision-dominated hydrodynamic limit have been considered here. For bosons
a necessary requirement is therefore a large scattering length and a
sufficiently large number-density
to ensure a  large cross-section for elastic collisions.
For fermions this requires, besides a large number-density, the simultaneous  
trapping of several hyperfine states, in order to allow for the interaction of  
the different fermionic species by elastic collisions, which would be  
forbidden
for a single species by the Pauli-principle. We have neglected
throughout the
spin-wave excitations, which can also occur in the latter systems, which is  
permitted because
they decouple from the density-waves by symmetry as long as the external  
potential is the same for all components. The collision-rates in degenerate  
Fermi-gases are suppressed by a Fermi-blocking factor compared to the  
classical collision rates \cite{18} and scale proportional to $(T/T_F)^2$.  
Therefore,
at least in the low-temperature domain, it is necessary to use atomic species  
with particularly large positive or negative s-wave scattering-lengths.  
Another limitation of our analysis is the neglection of mean-field effects of  
the interaction in comparison to the pressure term. This seems to be a rather  
good approximation for the experimentally realized trapped quantum-gases,  
which behave like ideal quantum-gases to a good approximation.
The most severe limitation of our calculations is certainly the neglect of  
dissipation.
The reason for this restriction (which is discussed further
in \cite{GWS} for bosons, and in \cite{BC,CG} for fermions)
lies not so much in the negligibility
of dissipative effects for the physically excited modes,
 but in the particular purpose we set out to achieve in this paper, namely to  
give an account of the  mode-spectrum in the whole temperature-domain
covered by the theory. This goal cannot be achieved so far with the inclusion  
of damping effects, but remains an interesting aim for future work. It seems  
clear that with the inclusion of damping the zero-frequency modes we have  
found will turn into purely over-damped modes. However,
as was exemplified for the Kohn-modes, the phenomenological theory behind
eq.(\ref{int}) permits to obtain also a result for the damping
of some modes (with the result that it is vanishing for the Kohn-modes)
whose hydrodynamic frequencies are only known
in the absence of dissipation \cite{KPS1,KPS2}. The requirement is
that the mode-frequencies without damping are also known in the collisionless  
limit. Then
an estimate of the collision-time can be used in eq.(\ref{int}) to obtain
an interpolation between the collision-dominated and the collisionless regime
including the damping due to the finite value of the collision-time \cite{KPS1,KPS2}.
In the past this estimate proved to be quite useful in the comparison of the  
experimental results for bosons  \cite{Ketterle},
and it may be hoped that further results along such lines not only for bosons  
but also for the rapidly developing experiments on trapped Fermi-gases
may be obtained in the near future.

\acknowledgements

This work has been supported by a project of the Hungarian Academy
of Sciences and the Deutsche Forschungsgemeinschaft under Grant No. 436 UNG 113/
144. A. Cs.
would like to acknowledge support by the Hungarian Academy of
Sciences under Grant No. AKP 98-20 2,2 and the
Hungarian National Scientific
Research Foundation under Grant Nos. OTKA F020094, T029552, T025866.
R. G.  wishes to acknowledge
support by the Deutsche
Forschungsgemeinschaft through the Sonderforschungbereich 237
"Unordnung and gro\ss e Fluktuationen".



\begin{references}
\bibitem{1} M.H.~Anderson \textit{et al}., Science \textbf{269}, 198
           (1995); K.B.~Davis \textit{et al}., Phys.~Rev.~Lett.
           \textbf{75}, 3969 (1995); C.C.~Bradley \textit{et al}.,
           \textit{ibid}. \textbf{78}, 985 (1997).
\bibitem{2} B.~DeMarco and D.S.~Jin, Science \textbf{285}, 1703 (1999)
\bibitem{3} M.-O.~Mewes, G.~Ferrari, F.~Schreck, A.~Sinatra, and C.~Salomon,
            physics/9909007 (September 6, 1999)
\bibitem{BK}L.P.~Kadanoff and G.~Baym, {\it Quantum Statistical Mechanics}
           (W.A.~Benjamin, N.Y., 1962), Ch.~6.
\bibitem{GWS} A.~Griffin, W.-C.~Wu, and S.~Stringari, Phys.~Rev.~Lett.
            \textbf{78}, 1838 (1997)
\bibitem{Kagan} Y. Kagan, E.L. Surkov and G. Shlyapnikov,
Phys. Rev. Lett. {\bf 789}, 2604 (1997)
\bibitem{BC} G.~M.~Bruun and Ch.~W.~Clark, cond-mat/9905263
\bibitem{A} M.~Amoruso, I.~Meccoli, A.~Minguzzi, and M.P.~Tosi,
            Eur.~Phys.~J.~D \textbf{7}, 441 (1999)
\bibitem{CG}  A.~Csord\'as and R.~Graham, cond-mat/0007049
\bibitem{VS} L.~Vichi and S.~Stringari, cond-mat/9905154
\bibitem{KPS1} G.M.~Kavoulakis, C.J.~Pethick, and H.~Smith,
Phys. Rev. Lett. {\bf 81}, 4036 (1998)
\bibitem{KPS2} G.M.~Kavoulakis, C.J.~Pethick, and H.~Smith,
Phys. Rev. A{\bf 57}, 2938 (1998)
\bibitem{APS} U.~Al~Khawaja, C.J.~Pethick and H.~Smith,
cond-mat/9908043
\bibitem{V}   L.~Vichi, cond-mat/0006305
\bibitem{GO} D. Gu\'ery-Odelin, F. Zambelli, J. Dalibard and S. Stringari,
cond-mat/9904409
\bibitem{LL} L.D Landau and E.M. Lifshitz, 'Fluid Mechanics' (Chapter 'Second  
viscosity'), Pergamon Press,
London (1959)
\bibitem{GriffinHyd} A. Griffin and E. Zaremba, Phys. Rev. A {\bf 56}, 4839
(1997); E. Zaremba, A. Griffin, and T. Nikuni, Phys. Rev. A{\bf 57}, 4695  
(1998); T. Nikuni and A. Griffin, J. Low Temp. Phys. {\bf 111}, 793 (1998);
 E. Zaremba, T. Nikuni, and A. Griffin, preprint cond-mat 9903029; M.  
Imamovic-Tomasovic and A. Griffin,
Phys. Rev. A {\bf 60}, 494 (1999)
\bibitem{CG2} A.~Csord\'as and R.~Graham, {\it Phys.~Rev. \bf A59}, 1477
           (1999).
\bibitem{L}  J.~Lighthill,`Waves in Fluids', Cambridge University Press,
             Cambridge 1978
\bibitem{TOP}D.S.~Jin, J.R.~Ensher, M.R.~Matthews, C.E.~Wieman,
           and E.A.~Cornell, Phys.~Rev.~Lett. \textbf{77}, 420 (1996);
           D.S.~Jin, M.R.~Matthews, J.R.~Ensher, C.E.~Wieman,
           and E.A.~Cornell, Phys.~Rev.~Lett. \textbf{78}, 764 (1997)
\bibitem{18} D.~Pines and P.~Nozi\`{e}res, \textit{The Theory of Quantum
            Liquids} (W.A.~Benjamin, New York, 1966), Vol.1
\bibitem{Ketterle} W. Ketterle, D.S. Durfee and D.M. Stamper-Kurn, in
"Proceedings of the International School of Physics - Enrico Fermi",
M. Inguscio and S. Stringari and C. E. Wieman editors, IOS Press 1999, p.67
\end{references}
\end{document}